\documentclass[11pt]{article}

\usepackage{color}
\usepackage{latexsym}
\usepackage{amssymb}
\usepackage{amsmath}
\usepackage{graphicx}
\usepackage{epstopdf}
\usepackage{subfig}
\usepackage{hhline}
\usepackage{float}

\newtheorem{Theorem}{Theorem}[part]

\newtheorem{Proposition}{Proposition}[part]

\newtheorem{Lemma}{Lemma}[part]

\newtheorem{Remark}{Remark}[part]

\def \R{\mathbb{R}}

\def \E{\mathbb{E}}
\def \F{\mathbb{F}}

\def \P{\mathbb{P}}

\def \Ac{{\cal A}}

\def \Cc{{\cal C}}

\def \Fc{{\cal F}}

\def \Lc{{\cal L}}

\def \Sc{{\cal S}}

\def \eps{\varepsilon}

\def \ep{\hbox{ }\hfill$\Box$}

\def\reff#1{{\rm(\ref{#1})}}

\def\beqs{\begin{eqnarray*}}
\def\enqs{\end{eqnarray*}}
\def\beq{\begin{eqnarray}}
\def\enq{\end{eqnarray}}

\begin{document}

\title{Optimal switching for  pairs trading rule: \\ a viscosity solutions approach}


\author{Minh-Man NGO
\\\small John von Neumann (JVN) Institute
\\\small Vietnam National University
\\\small Ho-Chi-Minh City,
\\\small man.ngo at jvn.edu.vn
\and 
Huy\^en PHAM 
 \\\small  Laboratoire de Probabilit\'es et
          \\\small  Mod\`eles Al\'eatoires, CNRS UMR 7599
             \\\small  Universit\'e Paris 7 Diderot,             
             \\\small   CREST-ENSAE, 
             \\\small  and JVN Institute
              \\\small  pham at math.univ-paris-diderot.fr 
             }
\maketitle

\begin{abstract}
This paper studies  the problem of determining the optimal cut-off for pairs trading rules. We consider two correlated assets whose spread is modelled by a mean-reverting process with stochastic volatility, and 
the optimal pair trading rule is formulated as an optimal switching problem between three regimes: flat position (no holding stocks), long one  short the other and short one  long the other. A fixed
commission cost is charged with each transaction. We use a viscosity solutions approach to prove the existence and the explicit characterization of cut-off points  via the resolution of quasi-algebraic equations.  We illustrate our results by numerical simulations.
\end{abstract}

\vspace{5mm}


\vspace{5mm}


\vspace{5mm}


 \newpage

\section{Introduction}
Pairs trading consists of taking simultaneously a long position in one of the assets $A$ and $B$, and 
a short position in the other,  in order to eliminate the market beta risk, and be exposed only to relative market movements determined by the spread. A brief history and discussion of pairs trading can be found in Ehrman \cite{ehr06}, Vidyamurthy  \cite{vidyamurthy2004pairs} and Elliott, Van der Hoek and Malcom  \cite{Elliott}. The main aim of this paper is to rationale mathematically these rules and find optimal  cutoffs, by means of a stochastic control approach.

 Pairs trading problem has been studied by stochastic control approach in the recent years.  Mudchanatongsuk, Primbs and Wong \cite{mudchanatongsuk2008optimal} consider  self-financing portfolio strategy for pairs trading,  
 model the log-relationship between a pair of stock prices by an Ornstein-Uhlenbeck process and use this to formulate a portfolio optimization and obtain the optimal solution to this control problem in closed
form via the corresponding Hamilton-Jacobi-Bellman (HJB) equation. They only allow positions that are short one stock and long the other, in equal dollar amounts. Tourin and Yan  \cite{tourin2013dynamic} study the same problem, but allow strategies with arbitrary amounts in each stock. On the other hand, instead of using self-financing strategies, one can focus on determining the optimal cut-offs, i.e. the boundaries of the trading regions in which one should trade when the spread lies in.  Such problem is closely related  to optimal buy-sell rule in trading mean reverting asset. Zhang and Zhang \cite{zhazha08} studied  optimal buy-sell rule, where they model the  underlying asset price by an Ornstein-Uhlenbeck process and consider an optimal trading rule determined by two regimes: buy and sell.  These regimes are defined by two threshold levels, and a fixed commission cost is charged with each transaction.  
They use classical verification approach to find the value function as solution to the associated HJB equations (quasi-variational inequalities),  and the optimal thresholds are obtained by smooth-fit technique. The same problem is studied in Kong's PhD thesis  \cite{kong2010stochastic}, but he considers  trading rules  with three aspects: buying, selling and shorting. Song and Zhang \cite{song2013optimal} use the same approach for determining optimal pairs trading thresholds, where they model the difference of the stock prices $A$ and $B$ by an Ornstein-Uhlenbeck process and consider an optimal pairs trading rule determined by two regimes: long $A$ short $B$ and flat position (no holding stocks).
Leung and Li \cite{leung2013optimal}  studied the optimal timing to open or close the position subject to transaction costs, and the effect of Stop-loss level under the Ornstein-Uhlenbeck (OU) model. They directly construct the
value functions instead of using  variational inequalities approach, by characterizing the value functions as the smallest concave majorant of reward function.

In this paper, we consider a pairs trading problem as in  Song and Zhang  \cite{song2013optimal}, but differ in our model setting and resolution method. We consider two correlated assets whose spread is modelled by a more general mean-reverting process with stochastic volatility, and the optimal pairs trading rule is based on optimal switching between three regimes: flat position (no holding stocks), long one  short the other and vice-versa. A fixed commission cost is charged with each transaction. We use a viscosity solutions approach to solve  our optimal switching problem.  Actually, by combining viscosity solutions approach,  smooth fit properties  and uniqueness result for viscosity solutions  proved in  Pham, Ly Vath and Zhou \cite{phalyvzho09},  we are able to derive directly  the structure of the switching regions, and thus the form of our value functions. 
This contrasts with the classical verification approach where the structure of the solution should be guessed ad-hoc, and one has to check that it  satisfies indeed the corresponding HJB equation,  which is not trivial in this context of optimal switching with more than two regimes.

The paper is organized as follows. We formulate in Section 2 the pairs trading as an optimal switching problem with three regimes. 
In Section 3, we state the system of variational inequalities satisfied by the value functions in the viscosity sense and the definition of pairs trading regimes. 
In Section 4,  we  state some useful properties on the switching regions, derive the form of value functions, and obtain optimal cutoff points by relying on the  smooth-fit properties  of value functions. In Section 5, we illustrate our results by numerical examples.

\section{Pair trading problem}

\setcounter{equation}{0} \setcounter{Assumption}{0}
\setcounter{Theorem}{0} \setcounter{Proposition}{0}
\setcounter{Corollary}{0} \setcounter{Lemma}{0}
\setcounter{Definition}{0} \setcounter{Remark}{0}

Let us consider the spread $X$ between two correlated assets, say $A$ and $B$ modelled by a mean-reverting process with boundaries  $\ell_-$ $\in$ $\{-\infty,0\}$, and $\ell_+$ $=$ $\infty$:
\beq \label{dynX}
dX_t &=&  \mu(L- X_t) dt + \sigma(X_t) dW_t,
\enq
where $W$ is a standard Brownian motion on $(\Omega,\Fc,\F=(\Fc_t)_{t\geq 0},\P)$, $\mu$ $>$ $0$ and $L$ $\geq$ $0$ are positive constants, $\sigma$ is a Lipschitz function on $(\ell_-,\ell_+)$, satisfying 
the nondegeneracy condition  $\sigma$ $>$ $0$. The SDE \reff{dynX} admits then a unique strong solution, given an initial condition $X_0$ $=$ $x$ $\in$ $(\ell_-,\ell_+)$,  denoted $X^x$. 
We assume that $\ell_+$ $=$ $\infty$ is a natural boundary, $\ell_-$ $=$ $-\infty$ is  a natural boundary, and  $\ell_-$ $=$ $0$ is non attainable. 
The main examples are the Ornstein-Uhlenbeck  (OU in short) process or the inhomogenous geometric Brownian motion (IGBM), as studied in detail   in the next sections.

Suppose that the investor starts with a flat position in both assets.  When the spread widens far from the equilibrium point,  she  naturally opens her trade by 
buying the underpriced asset, and selling the overpriced one. Next,  if the spread narrows,  she closes her trades, thus generating a profit.  
Such trading rules are quite popular in practice among hedge funds managers with cutoff values determined empirically  by descriptive statistics. 
The main aim of this paper is to rationale mathematically these rules and find optimal  cutoffs, by means of a stochastic control approach. More precisely, we 
formulate the pairs trading problem as an optimal switching problem with three regimes.  Let $\{-1,0,1\}$ be the set of  regimes where   
$i$ $=$ $0$ corresponds to a flat position (no stock holding),  $i$ $=$ $1$ denotes  a long position in the spread corresponding to a purchase of  $A$ and a sale of   $B$, while  $i$ 
$=$ $-1$ is a short position in $X$ (i.e.  sell $A$  and  buy $B$).  At any time, the investor can decide to open her trade by switching from regime $i$ $=$ $0$ to $i$ $=$ $-1$ (open to sell) or $i$ $=$ $1$ (open to buy). Moreover, when the investor is in a long ($i$ $=$ $1$) or short position ($i$ $=$ $-1$), she can decide to close her position by switching to regime $i$ $=$ $0$. 
We also assume that it is not possible for the investor to switch directly from regime $i$ $=$ $-1$ to $i$ $=$ $1$, and vice-versa, without first closing  
her position. 
The trading strategies of the investor are modelled by a switching control $\alpha$ $=$ $(\tau_n,\iota_n)_{n\geq 0}$ where $(\tau_n)_n$ is a nondecreasing sequence of stopping times representing the trading times, with $\tau_n$ $\rightarrow$ $\infty$ a.s. when $n$ goes to infinity, and $\iota_n$ valued in $\{-1,0,1\}$,  $\Fc_{\tau_n}$-measurable, represents the position regime decided at  $\tau_n$ until  the next trading time.  By misuse of notations, we denote by $\alpha_t$ the value of the regime at any time $t$:
\beqs
\alpha_t &=& \iota_0  {\bf 1}_{\{0 \leq t < \tau_0\}} +  \sum_{n \geq 0} \iota_n {\bf 1}_{\{\tau_n\leq t< \tau_{n+1}\}}, \;\;\; t \geq 0, 
\enqs
which also represents the inventory value in the spread at any time.  
We denote by $g_{ij}(x)$  the trading gain when switching from a position $i$ to $j$,  $i,j$ $\in$ $\{-1,0,1\}$, $j$ $\neq$ $i$, for  a spread value $x$.  The switching gain functions are given by: 
\beqs
g_{_{01}}(x) \; = \; g_{_{-10}}(x)  &=& - (x+ \eps) \\
g_{_{0-1}}(x) \; = \; g_{_{10}}(x) &=&  x-\eps,  
\enqs
where $\eps$ $>$ $0$ is a fixed transaction fee paid at each trading time.  Notice that we do not consider the functions $g_{_{-11}}$ and $g_{_{11}}$ since it is not possible to  switch 
from regime $i$ $=$ $-1$ to $i$ $=$ $1$ and  vice-versa.  By misuse of notations, we also set $g(x,i,j)$ $=$ $g_{_{ij}}(x)$. 

Given an initial spread value $X_0$ $=$ $x$, the expected reward over an infinite horizon associated to a switching trading strategy $\alpha$ $=$ 
$(\tau_n,\iota_n)_{n\geq 0}$  is given  by the gain functional: 
\beqs
J(x,\alpha) &=& \E \Big[  \sum_{n\geq 1} e^{-\rho \tau_n} g(X_{\tau_n}^x,\alpha_{\tau_n^-},\alpha_{\tau_n})   - 
 \lambda \int_0^{\infty}  e^{-\rho t} |\alpha_t| dt  \Big]. 
\enqs
The first (discrete sum) term  corresponds to  the (discounted with discount factor $\rho$ $>$ $0$) cumulated  gain of the investor by using pairs trading strategies,  while the last  integral term reduces the inventory risk, by penalizing with a factor $\lambda$ $\geq$ $0$, 
the  holding of  assets during the  trading time interval.

For $i$ $=$ $0,-1,1$,  let $v_i$ denote  the value functions with initial positions $i$ when maximizing over switching trading strategies the gain functional, that is
\beqs
v_i(x) &=& \sup_{\alpha\in\Ac_i} J(x,\alpha), \;\;\;\;\; x \in  (\ell_-,\infty), \; i =0,-1,1,
\enqs
where $\Ac_i$ denotes the set of switching controls  $\alpha$ $=$ $(\tau_n,\iota_n)_{n\geq 0}$ with initial position $\alpha_{0^-}$ $=$ $i$, i.e.  
$\tau_0$ $=$ $0$, $\iota_0$ $=$ $i$. The impossibility of switching directly from regime $i$ $=$ $\pm 1$ to $\mp 1$  is formalized by restricting the strategy of position $i= \pm 1$:  
if $\alpha \in \Ac_{_1}$ or $\alpha \in \Ac_{_{-1}}$ then $\iota_{_1}$ $=$ $0$ for ensuring that the investor has to close first her position before opening a new one.

\section{PDE characterization}

\setcounter{equation}{0} \setcounter{Assumption}{0}
\setcounter{Theorem}{0} \setcounter{Proposition}{0}
\setcounter{Corollary}{0} \setcounter{Lemma}{0}
\setcounter{Definition}{0} \setcounter{Remark}{0}

Throughout the paper, we denote by $\Lc$ the infinitesimal generator of the diffusion process $X$, i.e.
\beqs
\Lc \varphi(x)  &=& \mu(L- x) \varphi'(x) + \frac{1}{2} \sigma^2(x) \varphi''(x). 
\enqs 
The ordinary differential equation of second order
\beq \label{ode2}
\rho \phi  - \Lc \phi &=& 0,
\enq
has two linearly independent positive solutions. These solutions are uniquely determined (up to a multiplication), if we require one of them to be strictly 
increasing, and the other to be strictly decreasing. We shall denote by $\psi_+$ the increasing solution, and by $\psi_-$ the decreasing solution. They are called fundamental solutions of \reff{ode2}, and any other solution can be expressed as their linear combination.  
Since $\ell_+$ $=$ $\infty$ is a natural boundary, and $\ell_-$ $\in$  $\{-\infty,0\}$ is either a natural or non attainable boundary, we have: 
\beq \label{natural}
\psi_+(\infty) \; = \; \psi_-(\ell_-) \; = \; \infty, & & \psi_-(\infty) \; = \; 0. 
\enq
We shall also  assume that
\beq \label{limpsix}
\lim_{x\rightarrow \ell_-} \frac{x}{\psi_-(x)} \; = \; 0, & &  \lim_{x\rightarrow  \infty} \frac{x}{\psi_+(x)} \; = \; 0.
\enq

\vspace{2mm}

\noindent {\bf  Canonical examples} \\
Our two basic examples in finance for $X$ satisfying the above assumptions are
\begin{itemize}
\item  Ornstein-Uhlenbeck (OU) process:
\beq \label{OU}
dX_t &=& - \mu X_t dt + \sigma dW_t, 
\enq
with $\mu$, $\sigma$ positive constants.  In this case, $\ell_+$ $=$ $\infty$, $\ell_-$ $=$ $-\infty$ are natural boundaries,  the two fundamental solutions to \reff{ode2} are given by
\beqs
\psi_+(x)  =  \int_0^\infty t^{\frac{\rho}{\mu}-1}\exp\big(-\frac{t^2}{2} + \frac{\sqrt{2\mu}}{\sigma} x t\big) dt, & &  
\psi_-(x)  =  \int_0^\infty t^{\frac{\rho}{\mu}-1}\exp\big(-\frac{t^2}{2} -  \frac{\sqrt{2\mu}}{\sigma} x t\big) dt,
\enqs
and it is easily  checked that condition \reff{limpsix} is satisfied. 
 \item   Inhomogeneous Geometric Brownian Motion (IGBM):
\beq \label{inGBM}
dX_t =\mu(L- X_t) dt + \sigma X_t dW_t, \;\;\; X_{_0}>0, 
\enq
where $\mu$, $L$ and $\sigma$ are positive constants. In this case, $\ell_+$ $=$ $\infty$ is a natural boundary, $\ell_-$ $=$ $0$ is a non attainable boundary, and the two fundamental solutions to \reff{ode2} are given by
\beq \label{psi+-}
\psi_+(x)  =  x^{-a}U(a,b,\frac{c}{x}), & &  
\psi_-(x)  = x^{-a}M(a,b,\frac{c}{x}).
\enq
where 
\beq \label{igbmc}
a &=& \frac{\sqrt{\sigma^4+4(\mu+2\rho)\sigma^2+4\mu^2}-(2\mu+\sigma^2)}{2\sigma^2} \; > \; 0,\notag \\
b &=& \frac{2\mu}{\sigma^2}+2a+2, \;\;\; c \; = \; \frac{2\mu L}{\sigma^2},
\enq
and $M$ and $U$ are the confluent hypergeometric functions of the first and second kind.  Moreover, by the asymptotic property of the confluent hypergeometric functions (see \cite{abramowitz1972handbook}), 
the fundamental solutions $\psi_+$ and $\psi_-$ satisfy condition  \reff{limpsix}, and
\beq \label{limpsi2}
\psi_+(0^+) \; = \; \frac{1}{c^a}. 
\enq 
\end{itemize}

In this section, we state some general PDE characterization of the value functions by means of the dynamic programming approach. 
We first state a linear growth property and Lipschitz continuity  of the value functions.

\begin{Lemma} \label{lemgrowth}
There exists some positive constant $r$ (depending on $\sigma$) such that for a discount factor $\rho$ $>$ $r$, the value functions are finite on $\R$. In this case, we have
\beqs 
0 & \leq & v_{_0}(x) \; \leq \; C(1+|x|), \;\;\; \forall x \in  (\ell_-,\infty), \\ 
- \frac{\lambda}{\rho} & \leq & v_i(x) \; \leq \;  C(1+|x|), \;\;\; \forall x \in (\ell_-,\infty), \; i =1,-1,
\enqs
and
\beqs
|v_{_i}(x)-v_{_i}(y)|\; \leq \; C|x-y|, \;\;\; \forall x,y \in  (\ell_-,\infty), \; i =0,1,-1,
\enqs
for some positive constant $C$. 
\end{Lemma}
{\bf Proof.} The lower bound for $v_{_0}$ and $v_{i}$ are trivial by considering the strategies of doing nothing. 
Let us focus on the upper bound. First, by standard arguments using It\^o's formula and  Gronwall lemma, we have the following estimate on the diffusion $X$: there exists some positive  constant $r$, depending on 
the Lipschitz constant of $\sigma$, such that  
\beq 
\E |X_t^x| & \leq & C e^{r t} (1 + |x|), \;\;\; \forall t \geq 0, \label{estimX} \\
\E |X_t^x-X_t^y| & \leq & e^{r t} |x-y|  , \;\;\; \forall t \geq 0, \label{estimX2}
\enq
for some positive constant $C$ depending on $\rho$, $L$ and $\mu$. 
Next, for two successive trading times $\tau_n$ and $\sigma_n$ $=$ 
$\tau_{n+1}$ corresponding to a buy-and-sell or sell-and-buy strategy, we have: 
\beq
& & \E \Big[ e^{-\rho \tau_n} g(X_{\tau_n}^x,\alpha_{\tau_n^-},\alpha_{\tau_n})  + 
e^{-\rho \sigma_n} g(X_{\sigma_n}^x,\alpha_{\sigma_n^-},\alpha_{\sigma_n}) \Big]  \label{succes} \\
& \leq & \Big| \E\Big[ e^{-\rho\sigma_n} X_{\sigma_n}^x - e^{-\rho\tau_n} X_{\tau_n}^x \Big] \Big| \; \leq \;  \E \Big[ \int_{\tau_n}^{\sigma_n} e^{-\rho t} (\mu+\rho) |X_t^x| dt \Big]+ \E \Big[ \int_{\tau_n}^{\sigma_n} e^{-\rho t} \mu L dt \Big], \nonumber 
\enq
where the second inequality follows from It\^o's formula. When investor is staying in flat position $(i=0)$, in the first trading time investor can move to state $i=1$ or $i=-1$, and in the second trading time she has to back to state $i=0$. So that, the strategy when we stay in state $i=0$ can be expressed by the combination of infinite couples: \textit{buy-and-sell}, \textit{sell-and-buy}, for example: states $0 \rightarrow 1 \rightarrow 0 \rightarrow -1 \rightarrow 0 \rightarrow -1 \rightarrow 0 \rightarrow 1 \rightarrow 0 ...$ it means: buy-and-sell,  sell-and-buy, sell-and-buy, buy-and-sell,.... 
We deduce from \reff{succes}  that for any $\alpha$ $\in$ $\Ac_{_0}$,  
\beqs
J(x,\alpha) & \leq & \E \Big[ \int_0^\infty  e^{-\rho t} (\mu+\rho) |X_t^x| dt \Big]+\frac{\mu L}{\rho}. 
\enqs
Recalling that, when investor starts with a long or short position ($i$ $=$ $\pm 1$) she has to close first her position before opening a new one, so that 
for  $\alpha$ $\in$ $\Ac_{_1}$ or $\alpha$ $\in$ $\Ac_{_{-1}}$,
\beqs
J(x,\alpha) & \leq & |x|+\E \Big[ \int_{0}^{\tau_1} e^{-\rho t} (\mu+\rho) |X_t^x| dt \Big]+ \E \Big[ \int_{0}^{\tau_1} e^{-\rho t} \mu L dt \Big] \\
& & \; +   \E \Big[ \int_{\tau_2}^{\infty} e^{-\rho t} (\mu+\rho) |X_t^x| dt \Big]+ \E \Big[ \int_{\tau_2}^{\infty} e^{-\rho t} \mu L dt \Big] \\
& \leq &  |x|+ \E \Big[ \int_0^\infty  e^{-\rho t} (\mu+\rho) |X_t^x| dt \Big]+\frac{\mu L}{\rho},
\enqs
which proves the upper bound for $v_{i}$ by using the estimate \reff{estimX}. 
By the same argument, for two successive trading times $\tau_n$ and $\sigma_n$ $=$ 
$\tau_{n+1}$ corresponding to a buy-and-sell or sell-and-buy strategy, we have: 
\beqs
& & \E \Big[ e^{-\rho \tau_n} g(X_{\tau_n}^x,\alpha_{\tau_n^-},\alpha_{\tau_n})  +  
e^{-\rho \sigma_n} g(X_{\sigma_n}^x,\alpha_{\sigma_n^-},\alpha_{\sigma_n}) \\
& &  \;\;\;\;\; - \; e^{-\rho \tau_n} g(X_{\tau_n}^y,\alpha_{\tau_n^-},\alpha_{\tau_n})  - e^{-\rho \sigma_n} g(X_{\sigma_n}^y,\alpha_{\sigma_n^-},\alpha_{\sigma_n}) \Big] \\
& \leq & \Big| \E\Big[ e^{-\rho\sigma_n} X_{\sigma_n}^x - e^{-\rho\tau_n} X_{\tau_n}^x- e^{-\rho\sigma_n} X_{\sigma_n}^y + e^{-\rho\tau_n} X_{\tau_n}^y  \Big] \Big|\\
& \leq &  \E \Big[ \int_{\tau_n}^{\sigma_n} e^{-\rho t} (\mu+\rho) |X_t^x-X_t^y| dt \Big],
\enqs
where the second inequality follows from It\^o's formula.  We deduce that 
\beqs
|v_i(x)-v_i(y)|  &\leq&   \sup_{\alpha\in\Ac_i} |J(x,\alpha)-J(y,\alpha)| \\
& \leq & |x-y| +  \E \Big[ \int_0^\infty  e^{-\rho t} (\mu+\rho) |X_t^x-X_t^y| dt \Big],
\enqs
which proves the Lipschitz property for $v_{_i}, \; i=0,1,-1$ by using the estimate \reff{estimX2}. 
\ep

\vspace{3mm}

In the sequel, we fix a discount factor $\rho$ $>$ $r$ so that the value functions $v_i$ are well-defined and finite, and satisfy the linear growth and Lipschitz estimates of Lemma \ref{lemgrowth}. 
The dynamic programming equations satisfied by the value functions are thus given by a system of variational inequalities:
\beq
\min \big[ \rho v_{_0} - \Lc v_{_0} \; , \; v_{_0} - \max \big( v_{_1} + g_{_{01}}, v_{_{-1}} + g_{_{0-1}}\big) \big] &=& 0,  \;\;\; \mbox{ on } \; (\ell_-,\infty),  \label{pdev0} \\
\min \big[ \rho v_{_1}  - \Lc v_{_1} + \lambda \; , \;  v_{_1} - v_{_0} - g_{_{10}} \big] &=& 0,  \;\;\; \mbox{ on } \; (\ell_-,\infty),  \label{pdev1} \\
\min \big[ \rho v_{_{-1}}  - \Lc v_{_{-1}} + \lambda \; , \;  v_{_{-1}} - v_{_0} - g_{_{-10}} \big] &=& 0,  \;\;\; \mbox{ on } \; (\ell_-,\infty).  \label{pdev-1} 
\enq
Indeed, the equation for $v_0$ means that in regime $0$, the investor has the choice to stay in the flat position, or to open by a long or short position in the spread, while the equation for $v_i$, $i$ $=$ $\pm 1$, means that  in the regime $i$ $=$ $\pm 1$,  
she has first the obligation to close her position hence to switch to regime $0$ before opening a new position. 
By the same argument as in  \cite{pham2007smooth}, we know that the value functions $v_i, \; i=0,1,-1$ are viscosity solutions to the system 
(\ref{pdev0})-(\ref{pdev1})-(\ref{pdev-1}),  and satisfied the  smooth-fit  $C^1$ condition.

\vspace{1mm}

Let us  introduce the switching regions:     
\begin{itemize}
\item  Open-to-trade region from the flat position $i$ $=$ $0$:
 \beqs
 \Sc_{_0} & = & \Big\{ x \in (\ell_-,\infty) :   v_{_0}(x) =  \max \big( v_{_1} + g_{_{01}}, v_{_{-1}} + g_{_{0-1}}\big)(x) \Big\} \\
 &=&  \Sc_{_{01}} \cup \Sc_{_{0-1}},
 \enqs
where  $\Sc_{_{01}}$ is the  open-to-buy region, and  $\Sc_{_{0-1}}$ is the  open-to-sell region: 
\beqs
\Sc_{_{01}} &=& \Big\{ x \in (\ell_-,\infty) : v_{_0}(x) =  (v_{_1} + g_{_{01}})(x)  \Big\}, \\
\Sc_{_{0-1}} &=& \Big\{ x \in (\ell_-,\infty) : v_{_0}(x) =  (v_{_{-1}} + g_{_{0-1}})(x)  \Big\}.
\enqs 
\item  Sell-to-close region from the long position $i$ $=$ $1$:
\beqs
\Sc_{_1} &=& \Big\{ x \in (\ell_-,\infty) : v_{_1}(x) =  (v_{_0} + g_{_{10}})(x)  \Big\}.
\enqs 
\item  Buy-to-close region from the short position $i$ $=$ $-1$:
\beqs
\Sc_{_{-1}} &=& \Big\{ x \in (\ell_-,\infty) : v_{_{-1}}(x) =  (v_{_0} + g_{_{-10}})(x)  \Big\},
\enqs  
\end{itemize} 
 and the continuation regions, defined as the complement sets of the switching regions: 
\beqs
\Cc_{_0} &=& (\ell_-,\infty) \setminus\Sc_{_0} \; = \;   \Big\{ x \in (\ell_-,\infty) :   v_{_0}(x) >  \max \big( v_{_1} + g_{_{01}}, v_{_{-1}} + g_{_{0-1}}\big)(x) \Big\}, \\
\Cc_{_1} &=& (\ell_-,\infty) \setminus\Sc_{_1} \; = \;   \Big\{ x \in (\ell_-,\infty) :   v_{_1}(x) >   (v_{_0} + g_{_{10}})(x) \Big\}, \\
\Cc_{_{-1}} &=& (\ell_-,\infty) \setminus\Sc_{_{-1}} \; = \;   \Big\{ x \in (\ell_-,\infty) :   v_{_{-1}}(x) >   (v_{_0} + g_{_{-10}})(x) \Big\}. 
\enqs

\section{Solution}

\setcounter{equation}{0} \setcounter{Assumption}{0}
\setcounter{Theorem}{0} \setcounter{Proposition}{0}
\setcounter{Corollary}{0} \setcounter{Lemma}{0}
\setcounter{Definition}{0} \setcounter{Remark}{0}

In this section, we focus on the existence and structure of switching regions, and then we use the results on smooth fit property, uniqueness result for viscosity solutions of the value functions to derive the form of value functions in which the optimal cut-off points can be obtained  by solving smooth-fit condition equations.

\begin{Lemma} \label{leminclu}
\beqs
\Sc_{_{01}} \; \subset \;  \big(-\infty, \frac{\mu L -  \ell_{_0}}{\rho+\mu}\big]  \cap (\ell_-,\infty) ,  & & \Sc_{_{0-1}} \; \subset \; \big[ \frac{\mu L + \ell_{_0}}{\rho+\mu}, \infty\big),  \\
\Sc_{_1} \;  \subset \;  \big[ \frac{\mu L- \ell_{_1}}{\rho+\mu},\infty\big) \cap (\ell_-,\infty),  & & \Sc_{_{-1}} \; \subset \; \big(-\infty,  \frac{\mu L+ \ell_{_1}}{\rho+\mu}\big] \cap (\ell_-,\infty),
\enqs
where 
\beqs
0 \; < \; \ell_{_0} \; := \;  \lambda+\rho\eps, & & \ell_{_1} \; := \;   \lambda-\rho\eps   \;  \in \; (-\ell_{_0},\ell_{_0}). 
\enqs
\end{Lemma}
{\bf Proof.}
Let $\bar x$ $\in$ $\Sc_{_{01}}$, so that $v_{_0}(\bar x)$ $=$ $(v_{_1} + g_{_{01}})(\bar x)$.  By writing that $v_{_0}$ is a viscosity supersolution to: 
$\rho v_{_0}-\Lc v_{_0}$ $\geq$ $0$, we then get
\beq \label{ineg1}
\rho (v_{_1} + g_{_{01}})(\bar x) - \Lc (v_{_1} + g_{_{01}})(\bar x) & \geq & 0. 
\enq
Now, since  $g_{_{01}}+g_{_{10}}$ $=$ $-2\eps$ $<$ $0$, this implies that $\Sc_{_{01}}$ $\cap$ $\Sc_{_{1}}$ $=$ $\emptyset$,  so that $\bar x$ $\in$  
$\Cc_{_1}$. Since $v_{_1}$ satisfies the equation $\rho v_{_1} - \Lc v_{_1} + \lambda$ $=$ $0$ on $\Cc_{_1}$,  we then have  from \reff{ineg1}
\beqs
\rho  g_{_{01}}(\bar x)  - \Lc  g_{_{01}}(\bar x) - \lambda & \geq & 0.
\enqs
Recalling the expressions of $g_{_{01}}$ and $\Lc$, we thus obtain: 
$-\rho(\bar x+\eps) -  \mu\bar x - \lambda+L\mu$ $\geq$ $0$, which proves the inclusion result for $\Sc_{_{01}}$. 
Similar arguments show that if $\bar x$ $\in$  $\Sc_{_{0-1}}$ then
\beqs
\rho  g_{_{0-1}}(\bar x)  - \Lc  g_{_{0-1}}(\bar x) - \lambda & \geq & 0,
\enqs
which proves the inclusion result for $\Sc_{_{0-1}}$ after direct calculation. 

Similarly, if $\bar x$ $\in$ $\Sc_{_1}$ then $\bar x$ $\in$ $\Sc_{_{0-1}}$ or $\bar x$ $\in$ $\Cc_{_0}$:  if $\bar x$ $\in$ $\Sc_{_{0-1}}$,  we obviously have the inclusion result for $\Sc_{_1}$. On the other hand, if  $\bar x$ $\in$ $\Cc_{_0}$, using the viscosity supersolution property of $v_{_1}$, we have:
\beqs
\rho  g_{_{10}}(\bar x)  - \Lc  g_{_{10}}(\bar x) +  \lambda & \geq & 0,
\enqs
which yields the inclusion result for $\Sc_{_1}$.  By the same method, we shows  the inclusion result for $\Sc_{_{-1}}$.
\ep

\vspace{3mm}

We next  examine some sufficient conditions under which the switching regions are not empty.

\begin{Lemma} \label{lemempty1}
(1) The switching regions $\Sc_{_1}$ and $\Sc_{_{0-1}}$ are always not empty. 

\noindent (2) 
\begin{itemize}
\item[(i)] If $\ell_-$ $=$ $-\infty$, then  $\Sc_{_{-1}}$ is not empty  
\item[(ii)] If $\ell_-$ $=$ $0$, and  $\eps$ $<$ $\frac{\lambda}{\rho}$, then $\Sc_{_{-1}}$ $\neq$ $\emptyset$.
\end{itemize}

\noindent (3) If $\ell_-$ $=$ $-\infty$, then $\Sc_{_{01}}$ is not empty.  
\end{Lemma}
{\bf Proof.} (1) We argue by contradiction, and first assume that $\Sc_{_1}$ $=$ $\emptyset$. This means that once we are in the long position, it would be never optimal to close our position.  In other words, the value function $v_{_1}$ would be equal to $\hat V_{_1}$ given by
\beqs
\hat V_{_1}(x) &=& \E \Big[ - \lambda \int_0^{\infty} e^{-\rho t} dt \Big] \; = \; - \frac{\lambda}{\rho}. 
\enqs
Since $v_{_1}$ $\geq$ $v_{_0}+g_{_{10}}$, this would imply $v_{_0}(x)$ $\leq$ $- \frac{\lambda}{\rho} + \eps - x$, for all $x$ $\in$ $(\ell_-,\infty)$, which obviously contradicts the nonnegativity of the value function $v_{_0}$. 

Suppose now that $\Sc_{_{0-1}}$ $=$ $\emptyset$. Then, from the inclusion results for $\Sc_{_0}$ in Lemma \ref{leminclu}, this implies that the continuation region $\Cc_{_0}$ would contain at least the interval 
$(\frac{\mu L-\ell_{_0}}{\rho+\mu},\infty)$ $\cap$ $(\ell_-,\infty)$. In other words, we should have: 
$\rho v_{_0} - \Lc v_{_0}$ $=$ $0$ on    $(\frac{\mu L-\ell_{_0}}{\rho+\mu},\infty)$ $\cap$ $(\ell_-,\infty)$, and so  $v_{_0}$ should be in the form: 
\beqs
v_{_0}(x) &=&  C_+ \psi_+(x) + C_- \psi_-(x), \;\;\; \forall   x > \Big(\frac{\mu L-\ell_{_0}}{\rho+\mu} \Big) \vee \ell_-,
\enqs
for some constants $C_+$ and $C_-$. In view of the linear growth condition on $v_{_0}$ and condition \reff{limpsix} when $x$ goes to $\infty$, 
we must have $C_+$ $=$ $0$. 
On the other hand,  since $v_{_0}$ $\geq$ $v_{_{-1}}$ $+$ $g_{_{0-1}}$, and recalling the lower bound on $v_{_{-1}}$ in Lemma \ref{lemgrowth}, this would imply:
 \beqs 
 C_{-} \psi_-(x) & \geq  & - \frac{\lambda}{\rho} + x-\eps, \;\;\; \forall   x > \Big(\frac{\mu L-\ell_{_0}}{\rho+\mu} \Big) \vee \ell_-.   
\enqs
 By sending $x$ to $\infty$, and from \reff{natural}, we get the contradiction.

\vspace{2mm}

\noindent (2)  Suppose that $\Sc_{_{-1}}$ $=$ $\emptyset$. Then, a similar argument as in the case $\Sc_{_1}$ $=$ $\emptyset$, would imply that $v_{_0}(x)$ $\leq$ $- \frac{\lambda}{\rho} + \eps + x$, for all 
$x$ $\in$ $(\ell_-,\infty)$.  This immediately leads to a contradiction when $\ell_-$ $=$ $-\infty$ by sending $x$ to $-\infty$. When $\ell_-$ $=$ $0$, and under the condition that $\eps$ $<$ $\frac{\lambda}{\rho}$, 
we also get a contradiction to the non negativity of $v_{_0}$. 
 
\vspace{2mm}

\noindent (3)  Consider the case when $\ell_-$ $=$ $-\infty$, and let us argue by contradiction by assuming  that $\Sc_{_{01}}$ $=$ $\emptyset$.  Then, from the inclusion results for $\Sc_{_0}$ in Lemma \ref{leminclu}, this implies that the continuation region $\Cc_{_0}$ would contain at least the interval $(-\infty,\frac{\mu L+\ell_{_0}}{\rho+\mu})$. In other words, we should have: 
$\rho v_{_0} - \Lc v_{_0}$ $=$ $0$ on    $(-\infty,\frac{\mu L+\ell_{_0}}{\rho+\mu})$, and so  $v_{_0}$ should be in the form: 
\beqs
v_{_0}(x) &=&  C_+ \psi_+(x) + C_- \psi_-(x), \;\;\; \forall   x <  \frac{\mu L+\ell_{_0}}{\rho+\mu},
\enqs
for some constants $C_+$ and $C_-$. In view of the linear growth condition on $v_{_0}$ and condition \reff{limpsix} when $x$ goes to $-\infty$, 
we must have $C_-$ $=$ $0$.  On the other hand,  since $v_{_0}$ $\geq$ $v_{_{1}}$ $+$ $g_{_{01}}$, recalling the lower bound on $v_{_{1}}$ in Lemma \ref{lemgrowth}, this would imply:
\beqs
C_+ \psi_+(x) & \geq & - \frac{\lambda}{\rho} -(x+ \eps),  \;\;\; \forall   x <  \frac{\mu L+\ell_{_0}}{\rho+\mu}. 
\enqs
 By sending $x$ to $-\infty$, and from \reff{natural}, we get the contradiction.
\ep

\vspace{2mm}

\begin{Remark}
{\rm Lemma \ref{lemempty1} shows that $\Sc_{_1}$ is non empty. Furthermore, notice  that in the case where $\ell_-$ $=$ $0$, $\Sc_{_1}$ can be equal to 
the whole domain $(0,\infty)$, i.e. it is never optimal to stay in the long position regime.  Actually, from Lemma \ref{leminclu}, such extreme case may occur only if 
$\mu L - \ell_{_1}$ $\leq$ $0$, in which case, we would also get $\mu L - \ell_{_0}$ $<$ $0$, and thus $\Sc_{_{01}}$ $=$ $\emptyset$.  In that case, we are reduced to a 
problem with only two regimes $i$ $=$ $0$ and $i$ $=$ $-1$. 
\ep
}
\end{Remark}

\vspace{2mm}

The above Lemma \ref{lemempty1}  left open the question whether $\Sc_{_{-1}}$ is empty when $\ell_-$ $=$ $0$ and $\eps$ $\geq$ $\frac{\lambda}{\rho}$, and whether 
$\Sc_{_{01}}$ is empty or not when $\ell_-$ $=$ $0$. 
We examine this last issue in the next Lemma and the following remarks.

\begin{Lemma} \label{lemempty2} 
Let $X$ be governed by the Inhomogeneous Geometric Brownian motion in \reff{inGBM}, and set
\beqs \label{defK}
K_{0}(y) & := & (\frac{c}{y})^{-a}\frac{1}{U(a,b,\frac{c}{y})}(y-\eps+\frac{\lambda}{\rho})-(\frac{\lambda}{\rho}+\eps), \;\;\; y > 0,  \\ 
K_{-1}(y) & := & (\frac{c}{y})^{-a}\frac{1}{U(a,b,\frac{c}{y})}(y-\eps-\frac{\lambda}{\rho})+(\frac{\lambda}{\rho}-\eps), \;\;\; y > 0,
\enqs
where $a$, $b$ and $c$ are defined in \reff{igbmc}. If there exists $y$ $\in$ $(0,\frac{\mu L+\ell_{_0}}{\rho+\mu})$ (resp $y$ $>$ $0$) such that 
$K_{0}(y)$ (resp. $K_{-1}$) $>$ $0$,  then $\Sc_{_{01}}$ (resp. $\Sc_{_{-1}}$) is  not empty.  
\end{Lemma}
{\bf Proof.} Suppose that $\Sc_{_{01}}$ $=$ $\emptyset$. Then, from the inclusion results for $\Sc_{_0}$ in Lemma \ref{leminclu}, this implies that the continuation region 
$\Cc_{_0}$ would contain at least the interval $(0,\frac{\mu L+\ell_{_0}}{\rho+\mu})$. In other words, we should have: 
$\rho v_{_0} - \Lc v_{_0}$ $=$ $0$ on    $(0,\frac{\mu L+\ell_{_0}}{\rho+\mu})$, and so  $v_{_0}$ should be in the form: 
\beqs
v_{_0}(x) &=&  C_+ \psi_+(x) + C_- \psi_-(x), \;\;\; \forall  0 <  x <  \frac{\mu L+\ell_{_0}}{\rho+\mu},
\enqs
for some constants $C_+$ and $C_-$. From the bounds on $v_0$ in Lemma \ref{lemgrowth}, and \reff{natural}, we must have $C_-$ $=$ $0$. 

Next,  for $0<x\leq y$, let us consider the first passage time $\tau_y^x$ $:=$ $\inf\{t: X^{x}_t=y\}$ of the inhomogeneous Geometric Brownian motion. We know from \cite{zhao2009inhomogeneous} that
\beq \label{zhao1}
\E_x\big[ e^{-\rho \tau_y^x} \big] &=& \left(\frac{x}{y}\right)^{-a} \frac{U(a,b,\frac{c}{x})}{U(a,b,\frac{c}{y})}=\frac{\psi_+(x)}{\psi_+(y)}. 
\enq  
We denote by $\bar v_{_1}(x;y)$  the gain functional obtained from the strategy consisting in changing position from initial state $x$ and regime 
$i=1$, to the regime $i=0$ at the first time $X^{x}_t$ hits $y$ ($0<x \leq y$), and then following optimal decisions once in regime $i$ $=$ $0$: 
\beqs
\bar v_{_1}(x;y) &=&  \E [e^{-\rho \tau_y^x}(v_{_0}(y)+y-\eps)-\int_0^{\tau_y^x}\lambda e^{-\rho t}dt ], \;\;\; 0 < x \leq y. 
\enqs 
Since  $v_{_0}(y)$ $=$   $C_+ \psi_+(y)$,  for all $0<y< \frac{\mu L+\ell_{_0}}{\rho+\mu}$, and recalling (\ref{zhao1}) we have:
\beqs
\bar v_{_1}(x;y)  &=&  \E [e^{-\rho \tau_y^x}(C_+ \psi_+(y)+y-\eps)-\int_0^{\tau_y^x}\lambda e^{-\rho t}dt ] \\
 &=& \frac{\psi_+(x)}{\psi_+(y)}(C_+ \psi_+(y)+y-\eps+\frac{\lambda}{\rho})-\frac{\lambda}{\rho} \\
 &=& v_{_{0}}(x)+\frac{\psi_+(x)}{\psi_+(y)}(y-\eps+\frac{\lambda}{\rho})-\frac{\lambda}{\rho} , \; \; \; \; \; \; \forall 0<x \leq y<  \frac{\mu L+\ell_{_0}}{\rho+\mu}. 
\enqs
Now, by definition of $v_{_1}$, we have $v_{_1}(x)$ $\geq$ $\bar v_{_1}(x;y)$, so that:
\beqs \label{cont1}
v_{_1}(x) &\geq&  v_{_{0}}(x)+\frac{\psi_+(x)}{\psi_+(y)}(y-\eps+\frac{\lambda}{\rho})-\frac{\lambda}{\rho}, \; \; \; \forall 0<x \leq y< \frac{\mu L+\ell_{_0}}{\rho+\mu}.
\enqs
By sending $x$ to zero, and recalling \reff{psi+-} and \reff{limpsi2}, this yields
\beqs \label{abs1}
v_{_1}(0^+) &\geq&  v_{_0}(0^+) + K_0(y)+\eps, \; \; \; \forall 0<y< \frac{\mu L+\ell_{_0}}{\rho+\mu}.
\enqs
Therefore, under the condition that there exists $y$ $\in$ $(0,\frac{\mu L+\ell_{_0}}{\rho+\mu})$ such that 
$K(y)$ $>$ $0$, we would get:
\beqs
v_{_1}(0^+) &> &  v_{_0}(0^+) +\eps, 
\enqs
which is in contradiction with the fact that we have: $v_{_0}$ $\geq$ $v_{_1} + g_{_{01}}$, and so: $v_{_0}(0^+)$ $\geq$ $v_{_1}(0^+)$ $-$ $\eps$. 

\vspace{2mm}

Suppose that $\Sc_{_{-1}}$ $=$ $\emptyset$, in this case $v_{_{-1}}$ $=$ $-\lambda/\rho$.  
By the same argument as the above case,  we have
\beqs
v_{_0}(x) &\geq& \E[e^{-\rho \tau_y^x}(v_{_{-1}}(y)+y-\eps )] \; = \;  \E[e^{-\rho \tau_y^x}(-\frac{\lambda}{\rho}+y-\eps )] \notag \\
&=&  \big(-\frac{\lambda}{\rho}+y-\eps \big) \frac{\psi_+(x)}{\psi_+(y)}. 
\enqs
by \reff{zhao1}. 
By sending $x$ to zero, and recalling \reff{psi+-} and \reff{limpsi2}, we thus  have
\beq \label{condst1}
v_{_0}(0^+) &\geq& -\frac{\lambda}{\rho} + \eps + K_{-1}(y)  \;\;\;  y>0.
\enq
Therefore, under the condition that there exists $y$ $>$ $0$ such that 
$K_{-1}(y)$ $>$ $0$, we would get:
\beqs
v_{_0}(0^+) &> &  -\frac{\lambda}{\rho} +\eps, 
\enqs
which is in contradiction with the fact that we have: $v_{_{-1}}$ $\geq$ $v_{_0} + g_{_{-10}}$, and so: $ -\frac{\lambda}{\rho}$ $=$ $v_{_{-1}}(0^+)$ $\geq$ $v_{_0}(0^+)$ $-$ $\eps$. 
\ep

\vspace{2mm}

\begin{Remark}\label{remark41} {\rm The above Lemma \ref{lemempty2}  gives a sufficient condition in terms of the function $K_0$ and $K_{-1}$,  which ensures  that $\Sc_{_{01}}$ and $\Sc_{_{-1}}$ are not empty.  Let us discuss how it is satisfied. From the asymptotic property of the confluent hypergeometric functions, we have: $\lim_{z \rightarrow \infty} z^aU(a,b,z)=1$. Then by sending $L$ to infinity (recall that $c=\frac{2\mu L}{\sigma^2}$), and from the expression of $K_0$ and $K_{-1}$ in Lemma \ref{lemempty2}, we have:
\beqs
\lim_{L\rightarrow \infty} K_0(y) \; = \; \lim_{c \rightarrow \infty} K_0(y) &=&  y - 2 \eps \; = \;  \lim_{L\rightarrow \infty} K_{-1}(y).
\enqs
This implies that for $L$ large enough, one can choose $2 \eps < y < \frac{\mu L+\ell_{_0}}{\rho+\mu}$ so that $K_0(y)$ $>$ $0$. Notice also that 
$K_0$ is nondecreasing with $L$ as a consequence of the fact that $\frac{\partial}{\partial z}z^aU(a,b,z)=\frac{aU(a+1,b,z)(a-b+1)}{z}<0$. In practice, one can check by numerical method the condition $K_0(y)$ $>$ $0$ for $0 < y < \frac{\mu L+\ell_{_0}}{\rho+\mu}$.  
For example, with $\mu=0.8$,   $\sigma=0.5$ ,   $\rho=0.1$,   $\lambda=0.07$,   $\varepsilon=0.005$, and $L=3$, we have 
$\frac{\mu L+\ell_{_0}}{\rho+\mu}= 2.7450$, and $K_0(1)=0.9072>0$.  
Similarly, for $L$ large enough, one can find $y$ $>$ $2\eps $ such that $K_{-1}(y)$ $>$ $0$ ensuring that $\Sc_{_{-1}}$  is not empty. 
\ep
}
\end{Remark}

\vspace{2mm}

We are now able to describe the complete structure of the switching regions.

\begin{Proposition} \label{propcutoff} 
1) There exist finite cutoff levels  $\bar x_{_{01}}$, $\bar x_{_{0-1}}$, $\bar x_{_1}$, $\bar x_{_{-1}}$  such that
\beqs
\Sc_{_1} \; = \;  [\bar x_{_1},\infty) \cap (\ell_-,\infty), & & \Sc_{_{0-1}}  \; = \; [\bar x_{_{0-1}},\infty), \\
 \Sc_{_{-1}} \; = \; (\ell_-, -\bar x_{_{-1}}], & & \Sc_{_{01}} \; = \; (\ell_-, - \bar x_{_{01}}],
\enqs
and satisfying $\bar x_{_{0-1}}$ $\geq$ $\frac{\mu L+\ell_{_0}}{\rho+\mu}$, $\bar x_{_1}$ $\geq$ $\frac{\mu L -\ell_{_1}}{\rho+\mu}$, 
$-\bar x_{_{-1}}$ $\leq$ $\frac{\mu L+\ell_{_1}}{\rho+\mu}$,  $-\bar x_{_{01}}$ $\leq$ $\frac{\mu L-\ell_{_0}}{\rho+\mu}$. 
Moreover, $- \bar x_{_{01}}$ $<$ $\bar x_{_1}$, i.e.  $\Sc_{_{01}}$ $\cap$ $\Sc_{_1}$ $=$ $\emptyset$ and $ \bar x_{_{0-1}}$ $>$ $-\bar x_{_{-1}}$, i.e.  
$\Sc_{_{0-1}}$ $\cap$ $\Sc_{_{-1}}$ $=$ $\emptyset$.

\noindent 2)  We have $\bar x_{_1}$ $\leq$ $\bar x_{_{0-1}}$, and $-\bar x_{_{01}}$ $\leq$ $-\bar x_{_{-1}}$ i.e. the following  inclusions hold:
\beqs
\Sc_{_{0-1}} \; \subset \; \Sc_{_1}, & & \Sc_{_{01}} \; \subset \; \Sc_{_{-1}}. 
\enqs

\end{Proposition}
{\bf Proof.}  1) (i) We focus on the structure of the sets $\Sc_{_{01}}$ and $\Sc_{_{-1}}$, and consider first the case where they are not empty. 
Let us then set  $-\bar x_{_{01}}$ $=$ $\sup\Sc_{_{01}}$, which is finite 
since $\Sc_{_{01}}$ is not empty, and is included in $(\ell_-,\frac{\mu L-\ell_{_0}}{\rho+\mu}]$ by Lemma 
\ref{leminclu}.  Moreover, since $\Sc_{_{0-1}}$ is included in $[\frac{\mu L+\ell_{_0}}{\rho+\mu},\infty)$, it does not intersect with $(\ell_-,-\bar x_{_{01}})$, and so  $v_{_0}(x)$ $>$ 
$(v_{_{-1}}+ g_{_{0-1}})(x)$ for $x$ $<$  $-\bar x_{_{01}}$, i.e.  $(\ell_-,-\bar x_{_{01}})$ $\subset$ $\Sc_{_{01}}$ $\cup$ $\Cc_{_0}$.  From \reff{pdev0}, we deduce that $v_{_0}$ is a viscosity solution to 
\beq \label{pdev0-2}
\min \big[ \rho v_{_0} - \Lc v_{_0} \; , \; v_{_0} - v_{_1} - g_{_{01}} \big] &=& 0, \;\;\; \mbox{ on } \; (\ell_-,-\bar x_{_{01}}). 
\enq
Let us now prove that $\Sc_{_{01}}$ $=$ $(\ell_-,-\bar x_{_{01}}]$.  To this end, we consider the function $w_{_0}$ $=$ $v_{_1}+g_{_{01}}$ on $(\ell_-,-\bar x_{_{01}}]$. Let us check that 
$w_{_0}$ is a viscosity supersolution to
\beq \label{viscosurw0}
 \rho w_{_0} - \Lc w_{_0}  & \geq & 0  \;\;\; \mbox{ on } \; (\ell_-,-\bar x_{_{01}}).  
\enq
For this, take some point $\bar x$ $\in$ $(\ell_-,-\bar x_{_{01}})$, and some smooth test function $\varphi$ such that $\bar x$ is a local minimum of $w_{_0}-\varphi$. Then, $\bar x$ is a local minimum 
of $v_{_1}-(\varphi-g_{_{01}})$ by definition of $w_{_0}$.  By writing the viscosity supersolution property of $v_{_1}$ to: $\rho v_{_1} - \Lc v_{_1} + \lambda$ $\geq$ $0$, at $\bar x$ with the test function 
$\varphi-g_{_{01}}$, we get: 
\beqs
0  & \leq & \rho(\varphi-g_{_{01}})(\bar x) - \Lc(\varphi-g_{_{01}})(\bar x) + \lambda \\
&=& \rho \varphi(\bar x) - \Lc \varphi(\bar x)  + (\rho+\mu)(\bar x + \frac{\ell_{_0}- \mu L}{\rho+\mu}) \\
& \leq &   \rho \varphi(\bar x) - \Lc \varphi(\bar x),  
\enqs
since $\bar x$ $<$ $-\bar x_{_{01}}$ $\leq$ $\frac{\mu L-\ell_{_0}}{\rho+\mu}$.  This proves  the  viscosity supersolution property \reff{viscosurw0}, and actually, by recalling that $w_{_0}$ 
$=$ $v_{_1}+g_{_{01}}$,  
$w_{_0}$ is a viscosity solution to 
\beq \label{w0visco}
\min \big[ \rho w_{_0} - \Lc w_{_0} \; , \; w_{_0} - v_{_1} - g_{_{01}} \big] &=& 0, \;\;\; \mbox{ on } \; (\ell_-,-\bar x_{_{01}}).  
\enq
Moreover, since $-\bar x_{_{01}}$ lies in the closed set $\Sc_{_{01}}$, we have $w_{_0}(-\bar x_{_{01}})$ $=$ $(v_{_1}+g_{_{01}})(-\bar x_{_{01}})$ $=$ $v_{_0}(-\bar x_{_{01}})$.  By uniqueness 
of viscosity solutions to \reff{pdev0-2}, we deduce that $v_{_0}$ $=$ $w_{_0}$  on $(\ell_-,-\bar x_{_{01}}]$, i.e. $\Sc_{_{01}}$ $=$ $(\ell_-,-\bar x_{_{01}}]$. 
In the case where $\Sc_{_{01}}$ is empty, which may arise only when $\ell_-$ $=$ $0$ (recall Lemma \ref{lemempty1}), then it can still be written in the above form 
$(\ell_-,-\bar x_{_{01}}]$  by choosing  $-\bar x_{_{01}}$ $\leq$ $\ell_-$ $\wedge$  $(\frac{\mu L-\ell_{_0}}{\rho+\mu})$. 

By similar arguments, we show that when $\Sc_{_{-1}}$ is not empty, it should be in the form:   $\Sc_{_{-1}}$ $=$ $(\ell_-, -\bar x_{_{-1}}]$,  
for some $-\bar x_{_{-1}}$ $\leq$ $\frac{\mu L+\ell_{_1}}{\rho+\mu}$,  while when it is empty, which may arise only when $\ell_-$ $=$ $0$ (recall Lemma \ref{lemempty1}), it can be written also in this form by choosing $-\bar x_{_{-1}}$ $\leq$ $0$ $\wedge$ $(\frac{\mu L+\ell_{_1}}{\rho+\mu})$.

\vspace{1mm}

(ii)  We derive  similarly the structure of $\Sc_{_{0-1}}$ and $\Sc_{_1}$ which are already known to be non empty (recall Lemma \ref{lemempty1}): 
we set $\bar x_{_{0-1}}$ $=$ $\inf\Sc_{_{0-1}}$, which lies in $[\frac{\mu L+\ell_{_0}}{\rho+\mu},\infty)$  since $\Sc_{_{0-1}}$ is  included in 
$[\frac{\mu L+\ell_{_0}}{\rho+\mu},\infty)$ by Lemma \ref{leminclu}.  Then, we observe that $v_{_0}$ is a viscosity solution to 
\beq \label{pdev0-3}
\min \big[ \rho v_{_0} - \Lc v_{_0} \; , \; v_{_0} - v_{_{-1}} - g_{_{0-1}} \big] &=& 0, \;\;\; \mbox{ on } \; (\bar x_{_{0-1}},\infty). 
\enq
By considering the function $\tilde w_{_0}$ $=$ $v_{_{-1}}+g_{_{0-1}}$, we show  by the same arguments as in \reff{w0visco} that $\tilde w_{_0}$ is also a viscosity solution to \reff{pdev0-3} with 
boundary condition $\tilde w_{_0}(\bar x_{_{0-1}})$ $=$ $v_{_0}(\bar x_{_{0-1}})$. We conclude by uniqueness that $\tilde w_{_0}$ $=$ $v_{_0}$ on $[\bar x_{_{0-1}},\infty)$, i.e. 
$\Sc_{_{0-1}}$ $=$  $[\bar x_{_{0-1}},\infty)$. The same arguments show that $\Sc_{_1}$ is in the form stated in the Proposition. 

\vspace{1mm}

Moreover, from Lemma \ref{leminclu} we have : 
$\bar x_{_{0-1}}$ $\geq$ $\frac{\mu L+\ell_{_0}}{\rho+\mu}$ $>$ $\frac{\mu L+\ell_{_1}}{\rho+\mu}$ $\geq$  $-\bar x_{_{-1}}$ and  $\bar x_{_1}$ $\geq$ 
$\frac{\mu L-\ell_{_1}}{\rho+\mu}$ $>$ $\frac{\mu L-\ell_{_0}}{\rho+\mu}$ $\geq$  $-\bar x_{_{01}}$.

\vspace{3mm}

\noindent 2)  We only consider the case where $-\bar x_{_{-1}} < \bar x_{_{1}}$, since the inclusion result in this proposition is obviously obtained when  
$-\bar x_{_{-1}} \geq \bar x_{_{1}}$ from the above forms of the switching regions. 
Let us introduce the  function $U(x)=2v_{_0}(x)- (v_{_1}+v_{_{-1}})(x)$ on $[ -\bar x_{_{-1}},\bar x_{_{1}}]$. On $( -\bar x_{_{-1}},\bar x_{_{1}})$, we see that $v_{_1}$ and  $v_{_{-1}}$ are smooth $C^2$, and satisfy:
\beqs
\rho v_{_{1}}- \Lc v_{_{1}} +\lambda=0, & &  \rho v_{_{-1}}- \Lc v_{_{-1}} +\lambda=0,
\enqs
which combined with  the viscosity supersolution property of $v_{_0}$, gives
\beqs
\rho U- \Lc U \; = \; 2(\rho v_{_{0}}- \Lc v_{_{0}}) +2\lambda \; \geq \; 0 \ \ \  \text{on} \ \ \ (- \bar x_{_{-1}},\bar x_{_{1}}).
\enqs
At $x=\bar x_{_1}$ we have $v_{_1}(x)=v_{_0}(x)+x-\eps$ and $v_{_0}(x)\geq v_{_{-1}}(x)+x-\eps$ so that $2v_{_0}(x)\geq v_{_1}(x)+v_{_{-1}}(x)$, which means $U(\bar x_{_1})\geq 0$. By the same way, at $x=-\bar x_{_{-1}}$ we also have $2v_{_0}(x)\geq v_{_1}(x)+v_{_{-1}}(x)$, which means 
$U(-\bar x_{_{-1}})\geq 0$. By the  comparison principle, we deduce that
\beqs
2v_{_0}(x)\geq v_{_1}(x)+v_{_{-1}}(x) \ \ \ \text{on} \ \ \ [-\bar x_{_{-1}},\bar x_{_{1}}].
\enqs
Let us  assume on the contrary that $\bar x_{_1}$ $>$ $\bar x_{_{0-1}}$. We have $v_{_0}(\bar x_{_{0-1}})=v_{_{-1}}(\bar x_{_{0-1}})+\bar x_{_{0-1}}-\eps$ and $v_{_1}(\bar x_{_{0-1}})>v_{_0}(\bar x_{_{0-1}})+\bar x_{_{0-1}}-\eps$,  so that $(v_{_{-1}}+v_{_1})(\bar x_{_{0-1}})>2v_{_0}(\bar x_{_{0-1}})$, leading to a contradiction. By the same argument, it is impossible to have $-\bar x_{_{-1}}$ $<$ $-\bar x_{_{01}}$, which ends the proof.
\ep

 \vspace{5mm}

\begin{Remark}\label{remark42} {\rm 
Consider the situation where $\ell_-$ $=$ $0$.  We distinguish the following cases:
\begin{itemize}
\item[(i)] $\lambda$ $>$ $\rho \eps$. Then, we know from Lemma \ref{lemempty1} that $\Sc_{_{-1}}$ $\neq$ $\emptyset$. Moreover, for $L$ small enough, namely  $L$ $\leq$ 
$\ell_{_0}/\mu$, we see from Proposition \ref{propcutoff} that $-\bar x_{_{01}}$ $\leq$ $0$ and thus $\Sc_{_{01}}$ $=$ $\emptyset$. \\

\item[(ii)] $\lambda$ $\leq$ $\rho \eps$. Then $\ell_{_1}$ $\leq$ $0$, and for $L$ small enough namely, 
$L$ $\leq$ $-\ell_{_1}/\mu$, we see from Proposition \ref{propcutoff}  that $-\bar x_{_{-1}}$ $\leq$ $0$, and thus  $\Sc_{_{-1}}$ $=$ $\emptyset$ and $\Sc_{_{01}}$ $=$ $\emptyset$. 
\end{itemize}
\ep
}
\end{Remark}

\vspace{2mm}

The next result shows a symmetry property on the switching regions and value functions.

\begin{Proposition}\label{ldx} (Symmetry property)
In the case $\ell_-$ $=$ $-\infty$, and if $\sigma(x)$ is an even function and $L=0$, then $\bar x_{_{0-1}}=\bar x_{_{01}}$, $\bar x_{_{-1}}=\bar x_{_{1}}$ and 
\beqs
v_{_{-i}}(-x) & = & v_{_i}(x), \;\;\;\;\; x \in \R, \; i \in \{0,-1,1\}.
\enqs
\end{Proposition}
 {\bf Proof.}  Consider the process $Y^{x}_t=-X^x_t$,  which follows the dynamics: 
 \beqs
dY_t &=& -\mu Y_tdt + \sigma(Y_t)d\bar W_t,
\enqs
where $\bar W=-W$ is still a Brownian motion on the same probability measure and filtration of $W$, and we can see that $Y^{x}_t=X^{-x}_t$.  
 We consider the same optimal problem, but we use $Y_t$ instead of $X_t$, we denote 
 \beqs
J^Y(x,\alpha) &=& \E \Big[  \sum_{n\geq 1} e^{-\rho \tau_n} g(Y^x_{\tau_n},\alpha_{\tau_n^-},\alpha_{\tau_n})   - 
 \lambda \int_0^{\infty}  e^{-\rho t} |\alpha_t| dt  \Big],
\enqs
For $i$ $=$ $0,-1,1$,  let $v^{_Y}_i$ denote  the value functions with initial positions $i$ when maximizing over switching trading strategies the gain functional, that is
\beqs
v^{_Y}_{_i}(x) &=& \sup_{\alpha\in\Ac_i} J^Y(x,\alpha), \;\;\;\;\; x \in \R, \; i =0,-1,1. 
\enqs
For any $\alpha$ $\in$ $\Ac_i$,   we see that  $g(Y^x_{\tau_n},-\alpha_{\tau_n^-},-\alpha_{\tau_n})$ $=$ $g(X^x_{\tau_n},\alpha_{\tau_n^-},\alpha_{\tau_n})$, and so $J^Y(x,-\alpha)$ $=$ $J(x,\alpha)$.  Thus,  $v^{_Y}_{_{-i}}(x)$ $\geq$ $J^Y(x,-\alpha)$ $=$ $J(x,\alpha)$, and since $\alpha$ is arbitrary in $\Ac_i$, 
we get: $v^{_Y}_{_{-i}}(x)$  $\geq$ $v_{_i}(x)$.  By the same argument, we have $v_{_i}(x)\geq v^{_Y}_{_{-i}}(x)$, and so  $v_{_{-i}}^{_Y}$ $=$ $v_{_i}$, 
$i \in \{0,-1,1\}$. Moreover, recalling that  $Y^{x}_t=X^{-x}_t$, we have:
\beqs
v_{_{-i}}(-x)=v_{_{-i}}^{_Y}(x)=v_{_i}(x), \;\;\;\;\; x \in \R, \; i \in \{0,-1,1\}.
\enqs
In particular, we  $v_{_{-1}}(-\bar x_{_1})$ $=$ $v_{_1}(\bar x_{_1})=(v_{_0}+g_{_{10}})(\bar x_{_1})=(v_{_0}+g_{_{-10}})(-\bar x_{_1})$, so that  
$-\bar x_{_1} \in \Sc_{_{-1}}$. Moreover, since  $\bar x_{_1}$ $=$ $\inf \Sc_{_1}$, we notice that for all $r>0$, $\bar x_{_1}-r \not\in \Sc_{_{1}}$.  Thus, 
$v_{_{-1}}(-\bar x_{_1}+r)$ $=$  $v_{_1}(\bar x_{_1}-r)$ $>$ $(v_{_0}+g_{_{10}})(\bar x_{_1}-r)$ $=$ $(v_{_0}+g_{_{-10}})(-\bar x_{_1}+r)$,  for all $r$ $>$ $0$, which means that $-\bar x_{_1}$ $=$ 
$\sup \Sc_{_{-1}}$.  Recalling that $\sup \Sc_{_{-1}}$ $=$ $-\bar x_{_{-1}}$, this shows that $\bar x_{_1}$ $=$ $\bar x_{_{-1}}$.  By the same argument, we have $\bar x_{_{0-1}}=\bar x_{_{01}}$.
 \ep

\vspace{5mm}

\begin{figure}[H]
	\centering
\includegraphics[scale=0.6]{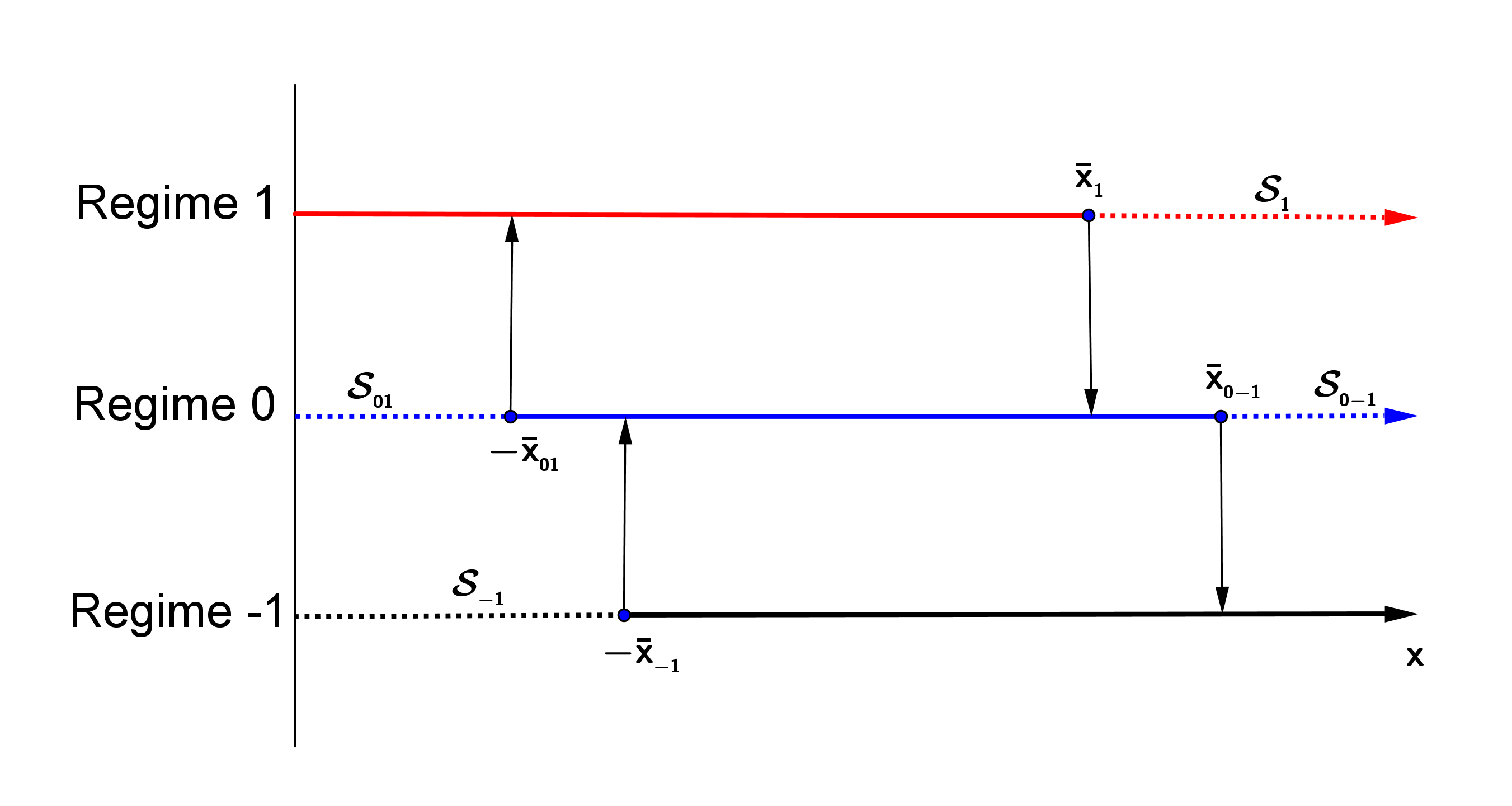}
\caption{\small \sl Regimes switching  regions in cases (1) and (2)(i). 
\label{figswitch1}} 
\end{figure}

\begin{figure}[H]
		\centering
	\includegraphics[scale=0.5]{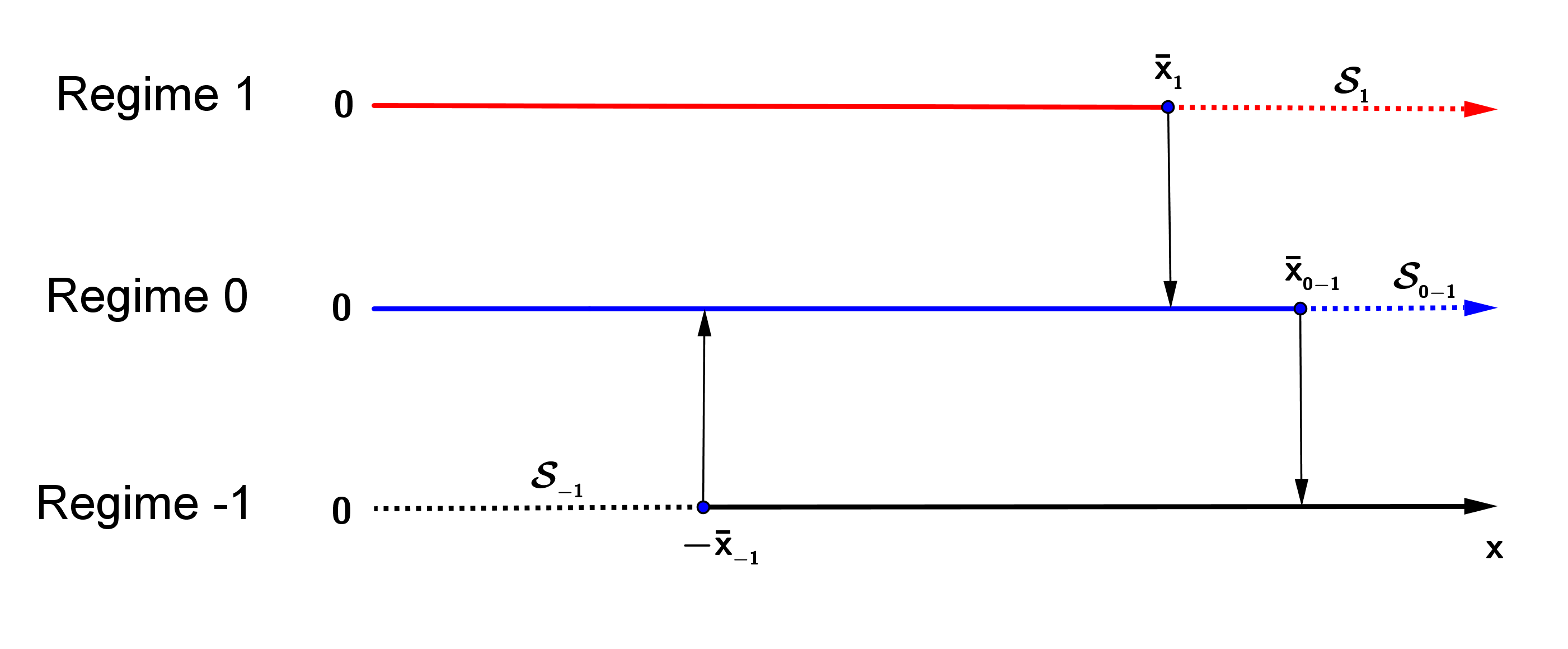}
	\caption{\small \sl Regimes switching  regions in case (2)(ii).
		\label{figswitch2}} 
\end{figure}

\begin{figure}[H]
	\centering
	\includegraphics[scale=0.5]{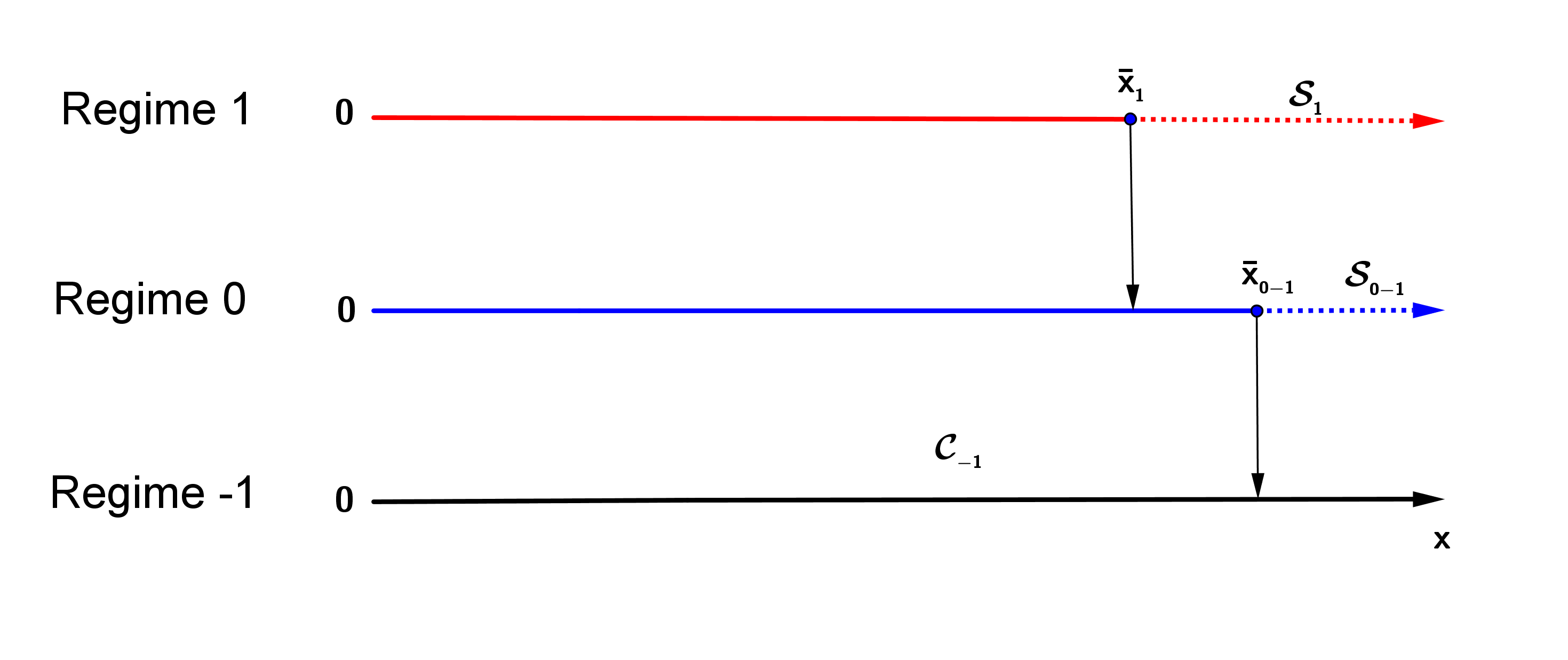}
	\caption{\small \sl Regimes switching  regions in case (2)(iii). 
		\label{figswitch3}} 
\end{figure}

 To sum up the above results, we have the following possible cases for the structure of the switching regions: 
\begin{itemize}
\item[{\bf (1)}] $\ell_-$ $=$ $-\infty$. In this case, the four switching regions $\Sc_{_1}$, $\Sc_{_{-1}}$, $\Sc_{_{01}}$ and $\Sc_{_{0-1}}$ are not empty in the form  
\beqs
\Sc_{_1} \; = \;  [\bar x_{_1},\infty), & & \Sc_{_{0-1}}  \; = \; [\bar x_{_{0-1}},\infty), \\
 \Sc_{_{-1}} \; = \; (-\infty, -\bar x_{_{-1}}], & & \Sc_{_{01}} \; = \; (-\infty, - \bar x_{_{01}}],
\enqs
and are  plotted in Figure \ref{figswitch1}. Moreover, when $L$ $=$ $0$ and $\sigma$ is an even function, $\Sc_{_1}$ $=$ $-\Sc_{_{-1}}$ and $\Sc_{_{01}}$ $=$ $-\Sc_{_{0-1}}$. 
\item[{\bf (2)}] $\ell_-$ $=$ $0$. In this case, the switching regions $\Sc_{_1}$ and $\Sc_{_{0-1}}$ are not empty, in the form
\beqs
\Sc_{_1} \; = \;  [\bar x_{_1},\infty) \cap (0,\infty), & & \Sc_{_{0-1}}  \; = \; [\bar x_{_{0-1}},\infty),
\enqs
for some $\bar x_{_{1}}$ $\in$ $\R$, and $\bar x_{_{0-1}}$ $>$ $0$ by Proposition \ref{propcutoff}. However,  $\Sc_{_{-1}}$ and $\Sc_{_{01}}$ may be empty or not. More precisely, we have the three following possibilities: 
\begin{itemize}
\item[(i)] $\Sc_{_{-1}}$ and $\Sc_{_{01}}$ are not empty in the form: 
\beqs
\Sc_{_{-1}} \; = \;  (0, -\bar x_{_{-1}}], & & \Sc_{_{01}} \; = \; (0, - \bar x_{_{01}}],
\enqs
for some $0$ $<$ $- \bar x_{_{01}}$  $\leq$ $-\bar x_{_{-1}}$ by Proposition \ref{propcutoff}. 
Such cases arises for example when $X$ is the IGBM \reff{inGBM} and for $L$ large enough, as showed in Lemma \ref{lemempty2} and 
Remark \ref{remark41}.    The visualization of this case is the same as Figure \ref{figswitch1}.
\item[(ii)] $\Sc_{_{-1}}$ is not empty in the form: $\Sc_{_{-1}}$  $=$ $(0, -\bar x_{_{-1}}]$ for some $\bar x_{_{-1}}$ $<$ $0$ by Proposition \ref{propcutoff}, 
and $\Sc_{_{01}}$ $=$ $\emptyset$. Such case arises  when $\lambda$ $>$ $\rho\eps$, and for $L$ $\leq$ $(\lambda+\rho\eps)/\mu$, see Remark \ref{remark42}(i). 
This is plotted in Figure \ref{figswitch2}. 
\item[(iii)]  Both $\Sc_{_{-1}}$  and $\Sc_{_{01}}$ are empty.  Such case arises 
when $\lambda$ $\leq$ $\rho\eps$, and for $L$  $\leq$ $(\rho\eps-\lambda)/\mu$, see Remark \ref{remark42}(ii). This is   plotted in Figure \ref{figswitch3}. 
Moreover, notice that in such case,  we must have  $\lambda$ $\leq$ $\rho\eps$ by Lemma \ref{lemempty1}(2)(ii), and so by Proposition \ref{propcutoff}, 
$\bar x_{_1}$ $\geq$  $\frac{\mu L -  \ell_{_1}}{\rho+\mu}$ $>$ $0$, i.e. $\Sc_{_1}$ $=$ $[\bar x_{_1},\infty)$.  

\end{itemize}
\end{itemize}

\vspace{2mm}

The next result provides  the explicit  solution to the optimal switching problem.

\begin{Theorem} \label{theo1}
$\bullet$ {\it Case (1): $\ell_-$ $=$ $\infty$}.  The value functions are given by
\beqs
v_{_0}(x) &=& \left\{ \begin{array}{cc}
				A_{_1} \psi_+(x) - \frac{\lambda}{\rho} + g_{_{01}}(x), & x \leq - \bar x_{_{01}}, \\
				A_{_0} \psi_+(x) + B_{_0} \psi_-(x), & - \bar x_{_{01}} < x < \bar x_{_{0-1}}, \\
				B_{_{-1}} \psi_-(x) - \frac{\lambda}{\rho} + g_{_{0-1}}(x), & x \geq  \bar x_{_{0-1}},
				\end{array}
				\right.  
\enqs
\beqs
v_{_1}(x) &=&  \left\{ \begin{array}{cc}
				A_{_1} \psi_+(x) - \frac{\lambda}{\rho}, & x <  \bar x_{_1}, \\
				v_{_0}(x) + g_{_{10}}(x), & x \geq   \bar x_{_1},
				\end{array}
				\right.
\enqs
\beqs
v_{_{-1}}(x) &=&  \left\{ \begin{array}{cc}
				v_{_0}(x) + g_{_{-10}}(x), & x  \leq   -\bar x_{_{-1}}, \\
				B_{_{-1}} \psi_-(x) - \frac{\lambda}{\rho}, & x >  -\bar x_{_{-1}},
				\end{array}
				\right.
\enqs
and the  constants $A_{_0}$, $B_{_0}$, $A_{_1}$, $B_{_{-1}}$, $\bar x_{_{01}}$,  $\bar x_{_{0-1}}$, 
$\bar x_{_1}$, $\bar x_{_{-1}}$ are determined by the smooth-fit conditions: 
\beqs
A_{_1} \psi_+(- \bar x_{_{01}}) - \frac{\lambda}{\rho} + g_{_{01}}(- \bar x_{_{01}}) &=& A_{_0} \psi_+(- \bar x_{_{01}}) + B_{_0} \psi_-(- \bar x_{_{01}}) \\
A_{_1} \psi_+'(- \bar x_{_{01}}) - 1 &=& A_{_0} \psi_+'(- \bar x_{_{01}}) + B_{_0} \psi_-'(- \bar x_{_{01}}) \\
B_{_{-1}} \psi_-(\bar x_{_{0-1}}) - \frac{\lambda}{\rho} + g_{_{0-1}}(\bar x_{_{0-1}}) &=& A_{_0} \psi_+( \bar x_{_{0-1}}) + B_{_0} \psi_-(\bar x_{_{0-1}})  \\
B_{_{-1}} \psi_-'(\bar x_{_{0-1}}) + 1 &=& A_{_0} \psi_+'( \bar x_{_{0-1}}) + B_{_0} \psi_-'(\bar x_{_{0-1}}) \\
A_{_1} \psi_+(\bar x_{_1}) - \frac{\lambda}{\rho} &=& A_{_0} \psi_+(\bar x_{_1}) + B_{_0} \psi_-(\bar x_{_1}) +  g_{_{10}}(\bar x_{_1}) \\
A_{_1} \psi_+'(\bar x_{_1}) &=& A_{_0} \psi_+'(\bar x_{_1}) + B_{_0} \psi_-'( \bar x_{_1}) + 1 \\
B_{_{-1}} \psi_-(-\bar x_{_{-1}}) - \frac{\lambda}{\rho} &=& A_{_0} \psi_+(-\bar x_{_{-1}}) + B_{_0} \psi_-(-\bar x_{_{-1}}) + g_{_{-10}}(-\bar x_{_{-1}}) \\
B_{_{-1}} \psi_-'(-\bar x_{_{-1}}) &=& A_{_0} \psi_+'(-\bar x_{_{-1}}) + B_{_0} \psi_-'(-\bar x_{_{-1}}) - 1. 
\enqs

\noindent $\bullet$ {\it Case (2)(i): $\ell_-$ $=$ $0$, and both $\Sc_{_{-1}}$ and $\Sc_{_{01}}$ are not empty}. The value functions  have  the same form  as Case (1) with  the state space 
domain  $(0,\infty)$.

\vspace{1mm}

\noindent  $\bullet$ {\it Case (2)(ii): $\ell_-$ $=$ $0$, $\Sc_{_{-1}}$ is not empty, and $\Sc_{_{01}}$ $=$ $\emptyset$}. 
 The value functions are given by
\beqs
v_{_0}(x) &=& \left\{ \begin{array}{cc}
	A_{_0} \psi_+(x) , & 0<  x < \bar x_{_{0-1}}, \\
	B_{_{-1}} \psi_-(x) - \frac{\lambda}{\rho} + g_{_{0-1}}(x), & x \geq  \bar x_{_{0-1}},
\end{array}
\right.  
\enqs
\beqs
v_{_{-1}}(x) &=&  \left\{ \begin{array}{cc}
	v_{_0}(x) + g_{_{-10}}(x), & 0< x  \leq   - \bar x_{_{-1}}, \\
	B_{_{-1}} \psi_-(x) - \frac{\lambda}{\rho}, & x >  - \bar x_{_{-1}},
\end{array}
\right.
\enqs
\beqs
v_{_1}(x) &=&  \left\{ \begin{array}{cc}
	A_{_1} \psi_+(x) - \frac{\lambda}{\rho}, & 0< x <  \max(\bar x_{_1},0), \\
	v_{_0}(x) + g_{_{10}}(x), & x \geq  \max(\bar x_{_1},0),
\end{array}
\right.
\enqs
and the  constants $A_{_0}$, $A_{_1}$, $B_{_{-1}}$,  $\bar x_{_{0-1}}$ $>$ $0$, 
$\bar x_{_1}$, $\bar x_{_{-1}}$ $<$ $0$ are determined by the smooth-fit conditions: 
\beqs
B_{_{-1}} \psi_-(\bar x_{_{0-1}}) - \frac{\lambda}{\rho} + g_{_{0-1}}(\bar x_{_{0-1}}) &=& A_{_0} \psi_+( \bar x_{_{0-1}})  \\
B_{_{-1}} \psi_-'(\bar x_{_{0-1}}) + 1 &=& A_{_0} \psi_+'( \bar x_{_{0-1}})\\
A_{_1} \psi_+(\bar x_{_1}) - \frac{\lambda}{\rho} &=& A_{_0} \psi_+(\bar x_{_1}) +  g_{_{10}}(\bar x_{_1}) \\
A_{_1} \psi_+'(\bar x_{_1}) &=& A_{_0} \psi_+'(\bar x_{_1}) +1 \\
B_{_{-1}} \psi_-(-\bar x_{_{-1}}) - \frac{\lambda}{\rho} &=& A_{_0} \psi_+(-\bar x_{_{-1}}) +  g_{_{-10}}(-\bar x_{_{-1}}) \\
B_{_{-1}} \psi_-'(-\bar x_{_{-1}}) &=& A_{_0} \psi_+'(-\bar x_{_{-1}}) - 1. 
\enqs

\vspace{1mm}

\noindent $\bullet$ {\it Case (2)(iii): $\ell_-$ $=$ $0$, and $\Sc_{_{-1}}$ $=$ $\Sc_{_{01}}$ $=$ $\emptyset$}.
The value functions are given by
\beqs
v_{_0}(x) &=& \left\{ \begin{array}{cc}
	A_{_0} \psi_+(x) , &  0< x < \bar x_{_{0-1}}, \\
	 - \frac{\lambda}{\rho} + g_{_{0-1}}(x), & x \geq  \bar x_{_{0-1}},
\end{array}
\right.  
\enqs
\beqs
v_{_1}(x) &=&  \left\{ \begin{array}{cc}
	A_{_1} \psi_+(x) - \frac{\lambda}{\rho}, & x <   \bar x_{_1}, \\
	v_{_0}(x) + g_{_{10}}(x), & x \geq   \bar x_{_1}, 
\end{array}
\right.
\enqs
\beqs
v_{_{-1}}=-\frac{\lambda}{\rho},
\enqs
and the  constants $A_{_0}$, $A_{_1}$, $\bar x_{_{0-1}}$ $>$ $0$,  $\bar x_{_1}$ $>$ $0$, are determined by the smooth-fit conditions: 
\beqs
 - \frac{\lambda}{\rho} + g_{_{0-1}}(\bar x_{_{0-1}}) &=& A_{_0} \psi_+( \bar x_{_{0-1}})  \\
1 &=& A_{_0} \psi_+'( \bar x_{_{0-1}})\\
A_{_1} \psi_+(\bar x_{_1}) - \frac{\lambda}{\rho} &=& A_{_0} \psi_+(\bar x_{_1}) +  g_{_{10}}(\bar x_{_1}) \\
A_{_1} \psi_+'(\bar x_{_1}) &=& A_{_0} \psi_+'(\bar x_{_1}) +1. 
\enqs
\end{Theorem}
{\bf Proof.}   We   consider only  case (1) and (2)(i) since the other cases are dealt with by similar arguments. We have 
$\Sc_{_{01}} \; = \; (\ell_-, - \bar x_{_{01}}]$, which means that $v_{_0}=v_{_1}+g_{_{01}}$ on $(\ell_-, - \bar x_{_{01}}]$. Moreover,  $v_{_1}$ is solution to 
$\rho v_{_{1}}- \Lc v_{_{1}} +\lambda=0$ on  $(\ell_-,\bar x_{_1})$, which combined with the bound in the Lemma \ref{lemgrowth}, shows that $v_{_1}$ should be in the form:  $v_{_1}$ $=$  $A_{_1} \psi_+ - \frac{\lambda}{\rho}$ on  $(\ell_-,\bar x_{_1})$. Since  $- \bar x_{_{01}}<\bar x_{_1}$, we deduce that 
$v_{_0}$ has the form expressed as: $A_{_1} \psi_+ - \frac{\lambda}{\rho} + g_{_{01}}$ on $(\ell_-, - \bar x_{_{01}}]$. In the same way, $v_{_{-1}}$ should have the form expressed as $B_{_{-1}} \psi_- - \frac{\lambda}{\rho}$ on  $(-\bar x_{_{-1}},\infty)$ and $v_{_0}$ has the form expressed as $B_{_{-1}} \psi_-  - \frac{\lambda}{\rho} + g_{_{0-1}}$ on $[\bar x_{_{0-1}}, \infty)$. 
We know that $v_{_0}$ is solution to $\rho v_{_{0}}- \Lc v_{_{0}} =0$ on $(- \bar x_{_{01}},\bar x_{_{0-1}})$ so that $v_{_0}$ should be in the form: 
$v_{_0}$ $=$ $A_{_0} \psi_+ + B_{_0} \psi_-$ on $(- \bar x_{_{01}},\bar x_{_{0-1}})$. We have $\Sc_{_1} \; = \;  [\bar x_{_1},\infty)$, which means that 
$v_{_1}$ $=$ $v_{_0}+g_{_{10}}$ on $[\bar x_{_1},\infty)$ and  $ \Sc_{_{-1}} \; = \; (\ell_-, -\bar x_{_{-1}}]$, which means that $v_{_{-1}}=v_{_0}+g_{_{-10}}$ on 
$(\ell_-, -\bar x_{_{-1}}]$.
From Proposition \ref{propcutoff} we know that $\bar x_{_1}$ $\leq$ $\bar x_{_{0-1}}$, and $-\bar x_{_{01}}$ $\leq$ $-\bar x_{_{-1}}$ and by the smooth-fit property of value function we obtain the above smooth-fit condition equations in which we can compute the cut-off points by solving these quasi-algebraic equations. 
\ep

\vspace{3mm}

\begin{Remark}		
{\rm  {\bf 1.} In  Case (1) and Case(2)(i) of  Theorem \ref{theo1}, the  smooth-fit conditions system is written as: 
\beq \label{sys1}
 \left[ \begin{array}{cccc}
\psi_+(-\bar{x}_{_{01}}) &0 & -\psi_+(-\bar{x}_{_{01}}) & -\psi_-(-\bar{x}_{_{01}})\\
0 & \psi_-(\bar{x}_{_{0-1}}) & -\psi_+(\bar{x}_{_{0-1}})& -\psi_-(\bar{x}_{_{0-1}})\\
\psi_+(\bar{x}_{_1})& 0 & -\psi_+(\bar{x}_{_1})& -\psi_-(\bar{x}_{_1})\\
0 & \psi_-(-\bar{x}_{_{-1}}) & -\psi_+(-\bar{x}_{_{-1}})&-\psi_-(-\bar{x}_{_{-1}})
\end{array} \right] \times
 \left[ \begin{array}{c}
A_{_1}\\B_{_{-1}}\\A_{_0}\\B_{_0}
\end{array} \right] = 
 \left[ \begin{array}{c}
\lambda \rho^{-1}-g_{_{01}}(-\bar{x}_{_{01}})\\ \lambda \rho^{-1}-g_{_{0-1}}(\bar{x}_{_{0-1}})\\ \lambda \rho^{-1}+g_{_{10}}(\bar{x}_{_{1}})\\ \lambda \rho^{-1}+g_{_{-10}}(-\bar{x}_{_{-1}})
\end{array} \right] \notag &\\
\enq
and
\beq \label{sys2}
 \left[ \begin{array}{cccc}
\psi_+^{'}(-\bar{x}_{_{01}}) &0 & -\psi_+^{'}(-\bar{x}_{_{01}}) & -\psi_-^{'}(-\bar{x}_{_{01}})\\
0 & \psi_-^{'}(\bar{x}_{_{0-1}}) & -\psi_+^{'}(\bar{x}_{_{0-1}})& -\psi_-^{'}(\bar{x}_{_{0-1}})\\
\psi_+^{'}(\bar{x}_{_1})& 0 & -\psi_+^{'}(\bar{x}_{_1})& -\psi_-^{'}(\bar{x}_{_1})\\
0 & \psi_-^{'}(-\bar{x}_{_{-1}}) & -\psi_+^{'}(-\bar{x}_{_{-1}})&-\psi_-^{'}(-\bar{x}_{_{-1}})
\end{array} \right] \times
 \left[ \begin{array}{c}
A_{_1}\\B_{_{-1}}\\A_{_0}\\B_{_0}
\end{array} \right] = 
 \left[ \begin{array}{c}
1\\-1\\1\\-1
\end{array} \right]. 
\enq
Denote by $M(\bar x_{_{01}},\bar x_{_{0-1}},\bar x_{_1},\bar x_{_{-1}})$ and $M_x(\bar x_{_{01}},\bar x_{_{0-1}},\bar x_{_1},\bar x_{_{-1}})$ the matrices: 
\beqs
M(\bar x_{_{01}},\bar x_{_{0-1}},\bar x_{_1},\bar x_{_{-1}}) &=& 
\left[ \begin{array}{cccc}
\psi_+(-\bar{x}_{_{01}}) &0 & 0 & -\psi_-(-\bar{x}_{_{01}})\\
0 & \psi_-(\bar{x}_{_{0-1}}) & -\psi_+(\bar{x}_{_{0-1}})& 0\\
\psi_+(\bar{x}_{_1})& 0 & 0& -\psi_-(\bar{x}_{_1})\\
0 & \psi_-(-\bar{x}_{_{-1}}) & -\psi_+(-\bar{x}_{_{-1}})&0
\end{array} \right], \\
M_x(\bar x_{_{01}},\bar x_{_{0-1}},\bar x_{_1},\bar x_{_{-1}}) &=& 
\left[ \begin{array}{cccc}
\psi_+^{'}(-\bar{x}_{_{01}}) &0 & 0 & -\psi_-^{'}(-\bar{x}_{_{01}})\\
0 & \psi_-^{'}(\bar{x}_{_{0-1}}) & -\psi_+^{'}(\bar{x}_{_{0-1}})& 0\\
\psi_+^{'}(\bar{x}_{_1})& 0 & 0& -\psi_-^{'}(\bar{x}_{_1})\\
0 & \psi_-^{'}(-\bar{x}_{_{-1}}) & -\psi_+^{'}(-\bar{x}_{_{-1}})&0
\end{array} \right]. 
\enqs
Once  $M(\bar x_{_{01}},\bar x_{_{0-1}},\bar x_{_1},\bar x_{_{-1}})$ and $M_x(\bar x_{_{01}},\bar x_{_{0-1}},\bar x_{_1},\bar x_{_{-1}})$ are nonsingular, 
straightforward computations from (\ref{sys1}) and (\ref{sys2})  lead to the following equation satisfied by  $\bar x_{_{01}}$,  $\bar x_{_{0-1}}$, 
$\bar x_{_1}$, $\bar x_{_{-1}}$:
\beqs \label{sys}
M_x(\bar x_{_{01}},\bar x_{_{0-1}},\bar x_{_1},\bar x_{_{-1}})^{-1} 
 \left[ \begin{array}{c}
1\\-1\\1\\-1
\end{array} \right] &=& 
M(\bar x_{_{01}},\bar x_{_{0-1}},\bar x_{_1},\bar x_{_{-1}})^{-1}  
 \left[ \begin{array}{c}
\lambda \rho^{-1}-g_{_{01}}(-\bar{x}_{_{01}})\\ \lambda \rho^{-1}-g_{_{0-1}}(\bar{x}_{_{0-1}})\\ \lambda \rho^{-1}+g_{_{10}}(\bar{x}_{_{1}})\\ \lambda \rho^{-1}+g_{_{-10}}(-\bar{x}_{_{-1}})
\end{array} \right]. 
\enqs
This system  can be separated into two independent systems:
\beq\label{quasi1}
 \left[ \begin{array}{cccc}
\psi_+^{'}(-\bar{x}_{_{01}})  & -\psi_-^{'}(-\bar{x}_{_{01}})\\
\psi_+^{'}(\bar{x}_{_1})& -\psi_-^{'}(\bar{x}_{_1})\\
\end{array} \right]^{-1} \times
 \left[ \begin{array}{c}
1\\1
\end{array} \right]=& \notag \\
\left[ \begin{array}{cccc}
\psi_+(-\bar{x}_{_{01}})  & -\psi_-(-\bar{x}_{_{01}})\\
\psi_+(\bar{x}_{_1}) & -\psi_-(\bar{x}_{_1})\\
\end{array} \right]^{-1} \times
 \left[ \begin{array}{c}
\lambda \rho^{-1}-g_{_{01}}(-\bar{x}_{_{01}})\\
 \lambda \rho^{-1}+g_{_{10}}(\bar{x}_{_{1}})\\
\end{array} \right]
\enq
and
\beq \label{quasi2}
 \left[ \begin{array}{cccc}
 \psi_-^{'}(\bar{x}_{_{0-1}}) & -\psi_+^{'}(\bar{x}_{_{0-1}})\\
 \psi_-^{'}(-\bar{x}_{_{-1}}) & -\psi_+^{'}(-\bar{x}_{_{-1}})
\end{array} \right]^{-1} \times
 \left[ \begin{array}{c}
-1\\-1
\end{array} \right]=& \notag\\
\left[ \begin{array}{cccc}
 \psi_-(\bar{x}_{_{0-1}}) & -\psi_+(\bar{x}_{_{0-1}})\\
 \psi_-(-\bar{x}_{_{-1}}) & -\psi_+(-\bar{x}_{_{-1}})
\end{array} \right]^{-1} \times
 \left[ \begin{array}{c}
 \lambda \rho^{-1}-g_{_{0-1}}(\bar{x}_{_{0-1}})\\
   \lambda \rho^{-1}+g_{_{-10}}(-\bar{x}_{_{-1}})
\end{array} \right]
\enq
We then obtain thresholds  $\bar x_{_{01}}$, $\bar x_{_{0-1}}$,  $\bar x_{_1}$, $\bar x_{_{-1}}$ by solving two quasi-algebraic system equations (\ref{quasi1}) and (\ref{quasi2}). Notice that for the examples of  OU  or IGBM process, the matrices 
$M(\bar x_{_{01}},\bar x_{_{0-1}},\bar x_{_1},\bar x_{_{-1}})$ and $M_x(\bar x_{_{01}},\bar x_{_{0-1}},\bar x_{_1},\bar x_{_{-1}})$ are nonsingular so that their inverses are well-defined. Indeed,  we have:  $\psi_+^{''}$ $>$ $0$ and $\psi_-^{''}$ $>$ $0$. This property is trivial for the case of OU process, while for  the case of IGBM: 
	\beqs
	\frac{d^2 \psi_+(x)}{dx^2} &=& \frac{d}{dx}\left( \frac{a}{x^{a+1}}(-U(a+1,b,\frac{c}{x})(a-b+1))\right)\\
	&=& \frac{a(a+1)}{x^{a+2}}U(a+2,b,\frac{c}{x})(a-b+1)(a-b+2) >0, \ \ \forall x>0.
	\enqs
	
	\beqs
	\frac{d \psi_-(x)}{dx} &=& -\frac{ax^{-a-2}(bxM(a,b,\frac{c}{x})+cM(a+1,b+1,\frac{c}{x}))}{b}, \ \ \forall x>0.
	\enqs
Thus, $\psi_-^{'}$ is strictly increasing since  $M(a,b,\frac{c}{x})$ is strictly decreasing, and so $\psi_-^{''}$ $>$ $0$. Moreover, we have:
\beq
& & \text{det} \;   \big[ M(\bar x_{_{01}},\bar x_{_{0-1}},\bar x_{_1},\bar x_{_{-1}}) \big]  \\
&=&   \left( \psi_-(\bar{x}_{_{-01}})\psi_+(\bar{x}_{_1})-
\psi_-(\bar{x}_{_1})\psi_+(\bar{x}_{_{-01}}) \right) \left(  \psi_-(\bar{x}_{_{0-1}})\psi_+(\bar{x}_{_{-1}})-
\psi_-(\bar{x}_{_{-1}})\psi_+(\bar{x}_{_{0-1}})\right).  \nonumber 
\enq
Recalling that   $- \bar x_{_{01}}$ $<$ $\bar x_{_1}$ and $ \bar x_{_{0-1}}$ $>$ $-\bar x_{_{-1}}$ (see Proposition \ref{propcutoff}), and since 
$\psi_+$ is a strictly increasing and  positive function, while  
$\psi_-$ is a strictly decreasing  positive function, we have: $ \psi_-(\bar{x}_{_{-01}})\psi_+(\bar{x}_{_1})-
\psi_-(\bar{x}_{_1})\psi_+(\bar{x}_{_{-01}})$ $>$ $0$ and $\psi_-(\bar{x}_{_{0-1}})\psi_+(\bar{x}_{_{-1}})-
\psi_-(\bar{x}_{_{-1}})\psi_+(\bar{x}_{_{0-1}}) $ $<$ $0$, which implies the non singularity  of the matrix 
$M(\bar x_{_{01}},\bar x_{_{0-1}},\bar x_{_1},\bar x_{_{-1}})$.   On the other hand, we have:
\beq
& & \text{det}  \big[ M_x(\bar x_{_{01}},\bar x_{_{0-1}},\bar x_{_1},\bar x_{_{-1}}) \big]  \\
&=&  \left(  \psi_-^{'}(\bar{x}_{_{-01}})\psi_+^{'}(\bar{x}_{_1})-
\psi_-^{'}(\bar{x}_{_1})\psi_+^{'}(\bar{x}_{_{-01}}) \right) \left(  \psi_-^{'}(\bar{x}_{_{0-1}})\psi_+^{'}(\bar{x}_{_{-1}})-
\psi_-^{'}(\bar{x}_{_{-1}})\psi_+^{'}(\bar{x}_{_{0-1}})\right).  \nonumber 
\enq
Since  $\psi_+'$ is a strictly increasing  positive function and  $\psi_-'$ is a strictly increasing function, with $\psi_-'$ $<$ $0$, we get: 
$ \psi_-^{'}(\bar{x}_{_{-01}})\psi_+^{'}(\bar{x}_{_1})- \psi_-^{'}(\bar{x}_{_1})\psi_+^{'}(\bar{x}_{_{-01}})$ $<$ $0$ and $\psi_-^{'}(\bar{x}_{_{0-1}})\psi_+^{'}(\bar{x}_{_{-1}})- \psi_-^{'}(\bar{x}_{_{-1}})\psi_+^{'}(\bar{x}_{_{0-1}}) $ $>$ $0$, which implies the non singularity of the matrix  
$M_x(\bar x_{_{01}},\bar x_{_{0-1}},\bar x_{_1},\bar x_{_{-1}})$.

\vspace{1mm}
 
\noindent {\bf 2.}  In  Case (2)(ii) of  Theorem \ref{theo1},  	we  obtain the thresholds   $\bar x_{_{0-1}}$ $>$ $0$, $\bar x_{_{-1}}$ $<$ $0$  from the smooth-fit conditions which lead to 
the quasi-algebraic system:
	\beq \label{sysx0t1xt12}
	\left[ \begin{array}{cccc}
		- \psi_-^{'}(\bar{x}_{_{0-1}}) & \psi_+^{'}(\bar{x}_{_{0-1}})\\
		
		-\psi_-^{'}(-\bar{x}_{_{-1}}) & \psi_+^{'}(-\bar{x}_{_{-1}})
	\end{array} \right]^{-1} \times
	\left[ \begin{array}{c}
		1\\1
	\end{array} \right]=& \notag \\
	\left[ \begin{array}{cccc}
		- \psi_-(\bar{x}_{_{0-1}}) & \psi_+(\bar{x}_{_{0-1}})\\
		
		-\psi_-(-\bar{x}_{_{-1}}) & \psi_+(-\bar{x}_{_{-1}})
	\end{array} \right]^{-1} \times
	\left[ \begin{array}{c}
		-\lambda \rho^{-1}+g_{_{0-1}}(\bar{x}_{_{0-1}})\\ -\lambda \rho^{-1}-g_{_{-10}}(-\bar{x}_{_{-1}})
	\end{array} \right].
	\enq
The non singularity of the matrix above is checked similarly as in case (1) and (2)(i) for the examples of the  OU  or IGBM process. 	
Note that  $\bar x_{_{0-1}}$, $\bar x_{_{-1}}$ are independent from $\bar x_{_1}$,  which is obtained from the equation: 
 	\beq \label{systx12}
	\left( -\lambda \rho^{-1}-g_{_{10}}(\bar{x}_{_{1}})\right)\psi_+^{'}(\bar{x}_{_{1}})+\psi_+(\bar{x}_{_{1}})=0.
	\enq
When $\bar x_{_1}$	 $\leq$ $0$, this means that $\Sc_{_1}$ $=$ $(0,\infty)$.

 \vspace{1mm}
 
 \noindent {\bf 3.}  In  Case (2)(iii) of  Theorem \ref{theo1}, the  threshold $\bar x_{_1}$ $>$ $0$  is obtained from the equation \reff{systx12}, while the threshold $\bar x_{_{0-1}}$ $>$ $0$ 
is derived from the smooth-fit condition leading to the  quasi-algebraic equation:
\beq \label{systx13}
\left( \lambda \rho^{-1}-g_{_{0-1}}(\bar{x}_{_{0-1}})\right)\psi_+^{'}(\bar{x}_{_{0-1}})+\psi_+(\bar{x}_{_{0-1}})=0.
\enq
 }
\ep
\end{Remark}

\section{Numerical examples}

\setcounter{equation}{0} \setcounter{Assumption}{0}
\setcounter{Theorem}{0} \setcounter{Proposition}{0}
\setcounter{Corollary}{0} \setcounter{Lemma}{0}
\setcounter{Definition}{0} \setcounter{Remark}{0}

In this part, we consider OU  process and IGBM as examples.

\begin{itemize}
\item[{\bf 1.}]  We first consider the example of the  Ornstein-Uhlenbeck process:
\end{itemize}
\beqs
dX_t &=&  -\mu X_t dt + \sigma dW_t, 
\enqs
with $\mu$, $\sigma$  positive constants.  In this case, the two fundamental solutions to \reff{ode2} are given by
\beqs
\psi_+(x)  =  \int_0^\infty t^{\frac{\rho}{\mu}-1}\exp\big(-\frac{t^2}{2} + \frac{\sqrt{2\mu}}{\sigma} x t\big) dt, & &  
\psi_-(x)  =  \int_0^\infty t^{\frac{\rho}{\mu}-1}\exp\big(-\frac{t^2}{2} -  \frac{\sqrt{2\mu}}{\sigma} x t\big) dt,
\enqs
and satisfy assumption \reff{limpsix}. We consider a numerical example with
the following specifications: :
$\mu=0.8$ ,   $\sigma=0.5$ ,   $\rho=0.1$ ,   $\lambda=0.07$ ,   $\varepsilon=0.005$ , $L=0$.

\begin{Remark} \label{remarkL}
{\rm  We can reduce the case of non zero long run mean $L \neq 0$ of the OU process  to the case of $L=0$ by considering process $Y_t=X_t-L$ as spread process, because in this case $\sigma$ is constant. Finally, we can see that, cutoff points translate along $L$, as illustrated  in figure \ref{fig:fig2}.
\ep
}
\end{Remark}

We recall some notations:\\
$\Sc_{_{01}} \; = \; (-\infty, - \bar x_{_{01}}]$ is the  open-to-buy region,\\
 $\Sc_{_{0-1}}  \; = \; [\bar x_{_{0-1}},\infty) $ is the  open-to-sell region,\\
$\Sc_{_1} \; = \;  [\bar x_{_1},\infty)$ is Sell-to-close region from the long position $i$ $=$ $1$,\\
$\Sc_{_{-1}} \; = \; (-\infty,-\bar x_{_{-1}}]$ is Buy-to-close region from the short position $i$ $=$ $-1$.

We solve the  two systems \reff{quasi1} and \reff{quasi2} which give
\beqs
\bar x_{_{01}}= 0.2094, \;  \bar x_{_1}=0.0483, \;  \bar x_{_{-1}}= 0.0483,  \; \bar x_{_{0-1}}=   0.2094,
\enqs
and confirm the symmetry property in Proposition \ref{ldx}.

\begin{figure}[H]
\includegraphics[scale=0.5]{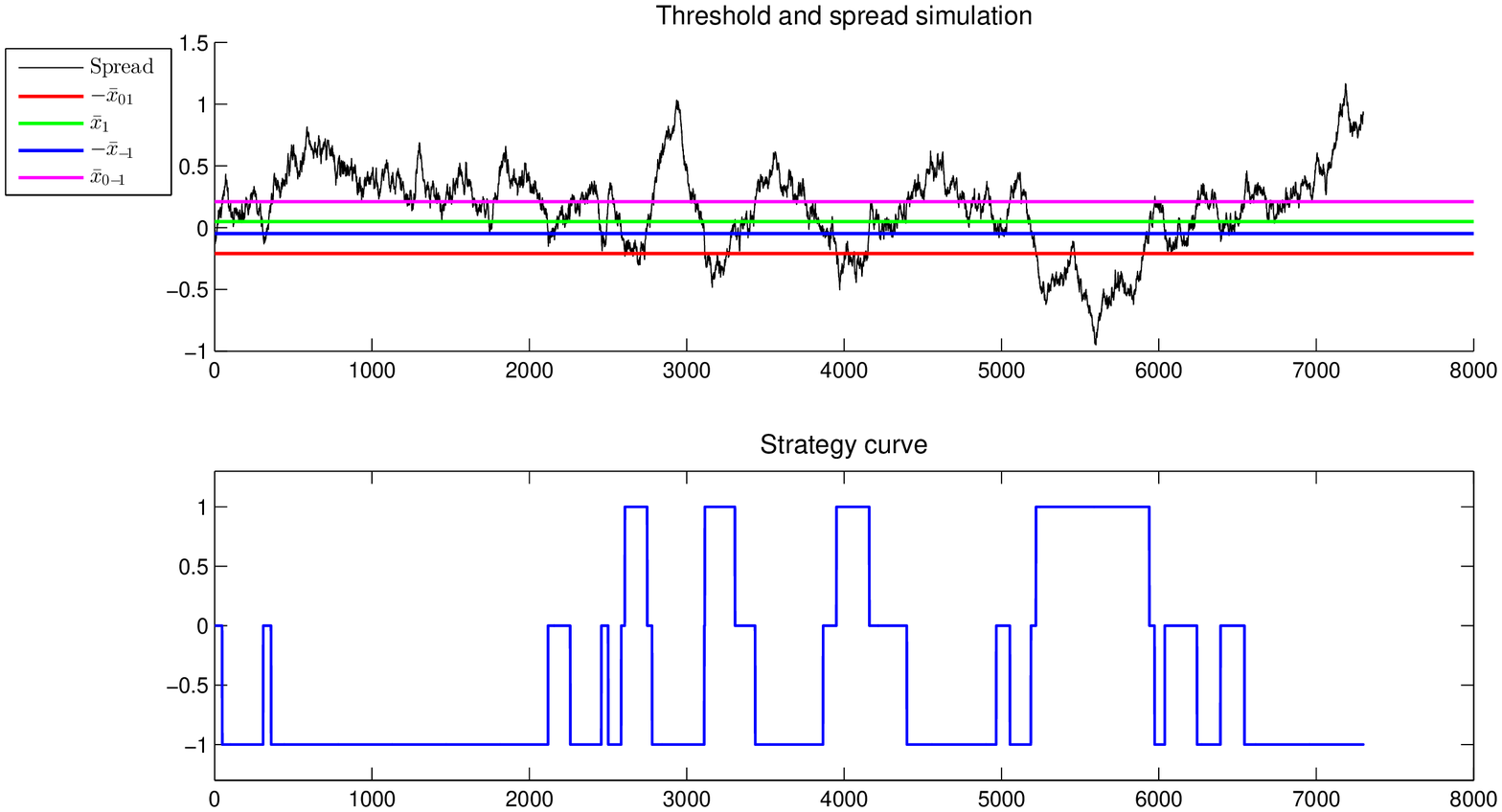}
\centering
\caption{\small \sl Simulation of trading strategies  
\label{fig:fig1}} 
\end{figure}
\begin{figure}[H]
	\includegraphics[scale=0.6]{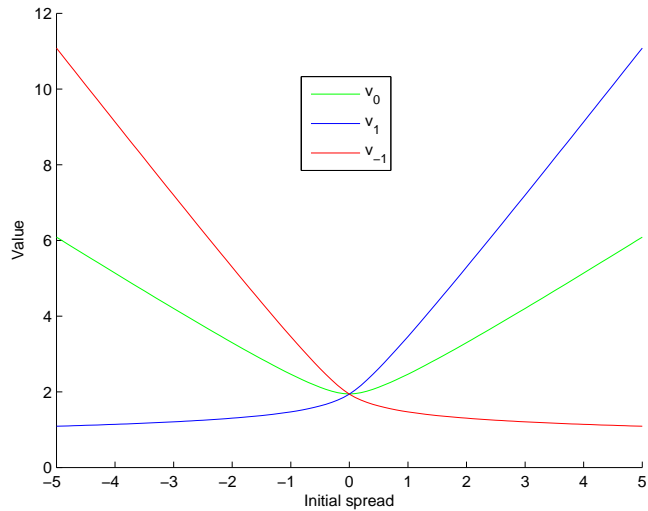}
	\centering
	\caption{\small \sl Value functions
		\label{fig:figvalue0}} 
\end{figure}
In figure \ref{fig:figvalue0}, we see the symmetry property of value functions as showed in Proposition \ref{ldx}.  
Moreover, we can see that $v_{_1}$ is a non decreasing function  while  $v_{_{-1}}$ is non increasing.

The next figure shows  the dependence of cut-off point on parameters
\begin{figure}[H]
\includegraphics[scale=0.6]{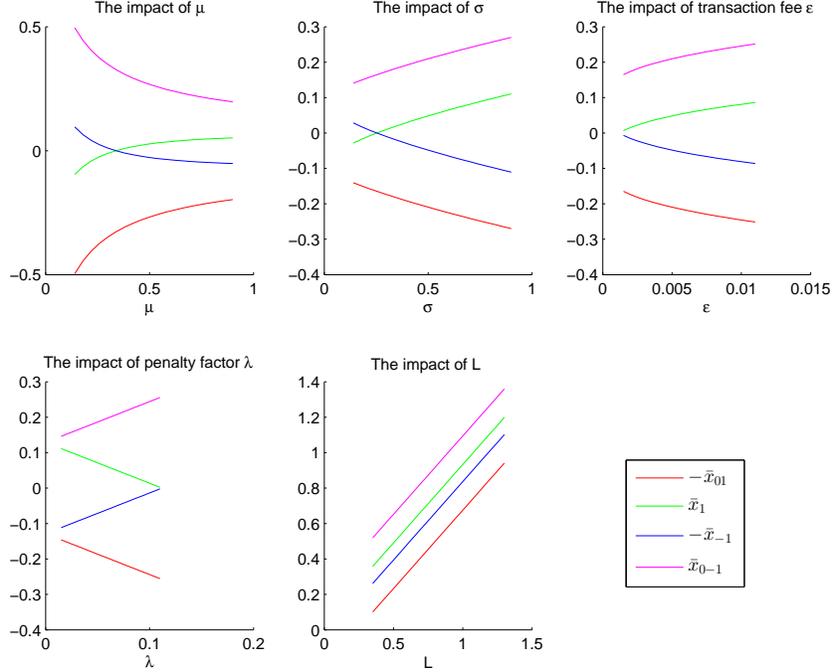}
\centering
\caption{\small \sl The dependence of cut-off point on parameters 
\label{fig:fig2}} 
\end{figure}

In figure \ref{fig:fig2}, $\mu$ measures the speed of mean reversion and we see that the length of intervals $\Sc_{_{01}}$, $\Sc_{_{0-1}}$ increases and the length of intervals
$\Sc_{_1}$, $\Sc_{_{-1}}$ decreases as $\mu$ gets bigger. The length of intervals $\Sc_{_{01}}$, $\Sc_{_{0-1}}$, $\Sc_{_1}$, and $\Sc_{_{-1}}$ decreases as volatility $\sigma$ gets bigger. $L$ is the long run mean€™, to which the process tends to revert, and we see that the cutoff points translate along $L$. We now look at the parameters that does not affect on the dynamic of spread: the length of intervals $\Sc_{_{01}}$, $\Sc_{_{0-1}}$, $\Sc_{_1}$, and $\Sc_{_{-1}}$ decreases as the transaction fee $\eps$ gets bigger. Finally, the length of intervals $\Sc_{_{01}}$, $\Sc_{_{0-1}}$ decreases and the length of intervals
$\Sc_{_1}$, $\Sc_{_{-1}}$ increases as the penalty factor $\lambda$ gets larger, which means  that the holding time in flat position $i=0$ is longer and the opportunity to enter the flat position from the other position is bigger as the penalty factor $\lambda$ is increasing.

\begin{itemize}
\item[{\bf 2.}] We now consider the example of Inhomogeneous Geometric Brownian Motions which has stochastic volatility, see more details  in Zhao \cite{zhao2009inhomogeneous} :
\end{itemize}
\beqs
dX_t =\mu(L- X_t) dt + \sigma X_t dW_t, \;\;\; X_{_0}>0,
\enqs
where $\mu$, $L$ and $\sigma$ are positive constants. Recall that in this case, the two fundamental solutions to \reff{ode2} are given by
\beqs
\psi_+(x)  =  x^{-a}U(a,b,\frac{c}{x}), & &  
\psi_-(x)  = x^{-a}M(a,b,\frac{c}{x}),
\enqs
where 
\beqs
a &=& \frac{\sqrt{\sigma^4+4(\mu+2\rho)\sigma^2+4\mu^2}-(2\mu+\sigma^2)}{2\sigma^2}>0, \\
b &=& \frac{2\mu}{\sigma^2}+2a+2, \;\;\;\;\; c=\frac{2\mu L}{\sigma^2},
\enqs
$M$ and $U$ are the confluent hypergeometric functions of the first and second kind. We can easily check that $\psi_-$ is a monotone decreasing function, while
\beqs
\frac{d \psi_+(x)}{dx} &=& \frac{a}{x^{a+1}}(-U(a+1,b,\frac{c}{x})(a-b+1))>0, \ \ \forall x>0,
\enqs
so that   $\psi_+$ is  a monotone increasing  function. Moreover, by the asymptotic property of the confluent hypergeometric functions (cf.\cite{abramowitz1972handbook}), the fundamental solutions $\psi_+$ and $\psi_-$ satisfy  the  condition \reff{limpsix}. 

\vspace{1mm}

\noindent $\bullet$ {\it Case (2)(i)}: Both $\Sc_{_{-1}}$ and $\Sc_{_{01}}$ are not empty. 
Let us  consider a numerical example with the following specifications: :
$\mu=0.8$ ,   $\sigma=0.5$ ,   $\rho=0.1$ ,   $\lambda=0.07$ ,   $\varepsilon=0.005$ , and we set $L=10$. Note that, in this case the condition in Lemma \ref{lemempty2} is satisfied,
and we solve the two systems (\ref{quasi1}) and (\ref{quasi2}) which give
\beqs
\bar x_{_{01}}= - 8.2777,   \; \bar x_{_1}= 9.3701, \;  \bar x_{_{-1}} = - 8.4283, \;  \bar x_{_{0-1}}=    9.5336.
\enqs
\begin{figure}[H]
	\includegraphics[scale=0.6]{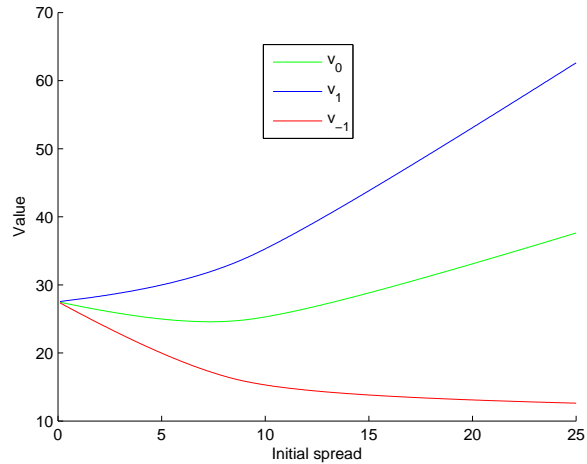}
	\centering
	\caption{\small \sl Value functions
		\label{fig:figvalue1}} 
\end{figure}
In the figure \ref{fig:figvalue1},  we can see that $v_{_1}$ is non decreasing while  $v_{_{-1}}$ is non increasing. Moreover, $v_{_1}$ is always larger than $v_{_0}$, and $v_{_{-1}}$.

The next figure \ref{fig:fig3} shows the dependence of cut-off points on parameters (Note that the condition in Lemma \ref{lemempty2} is satisfied for all parameters in this figure).

\begin{figure}[H]
\includegraphics[scale=0.65]{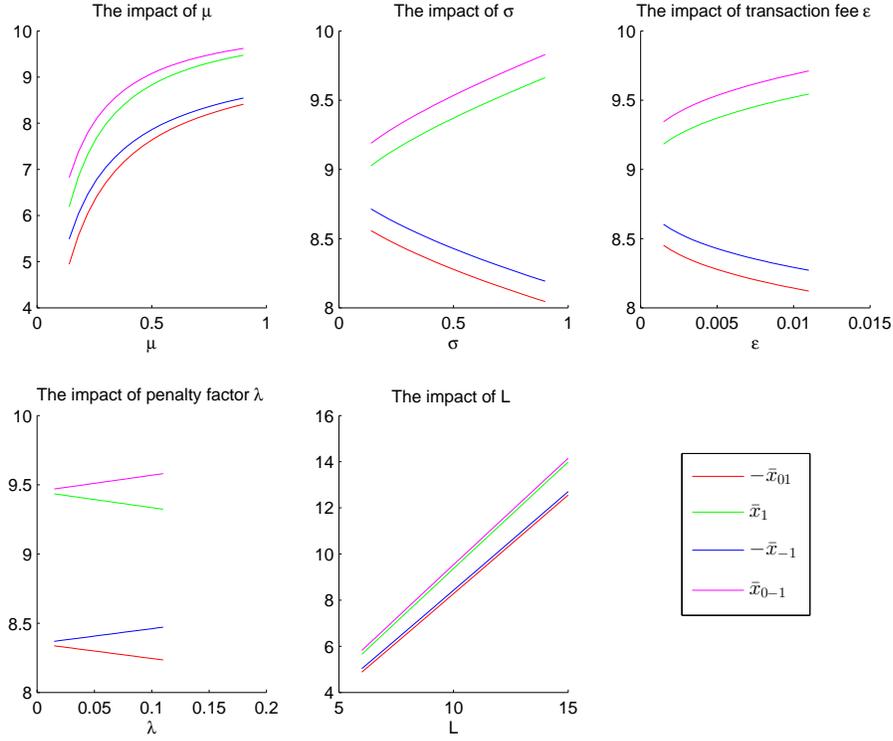}
\centering
\caption{\small \sl The dependence of cut-off point on parameters 
\label{fig:fig3}} 
\end{figure}

We can make the same comments as in the case of the OU process, except  for the dependence with respect to the long run mean $L$. 
Actually, we see that when $L$ increases,  the moving of cutoff points is no more translational due to the non constant volatility. 

\vspace{3mm}

\noindent $\bullet$ {\it Case (2)(ii)}: $\Sc_{_{01}}$ is empty. 
Let us  consider a numerical example with the following specifications: :
$\mu=0.8$,   $\sigma=0.3$,   $\rho=0.1$,   $\lambda= 0.35$,   $\varepsilon=0.55$,  and $L=0.5$. We solve the two systems (\ref{sysx0t1xt12}) and (\ref{systx12}) which give
\beqs 
\bar x_{_1}= 0.1187, \; \bar x_{_{-1}} = -0.8349, \;  \bar x_{_{0-1}}=     2.7504.
\enqs

\noindent $\bullet$ {\it Case (2)(iii)}: Both $\Sc_{_{-1}}$ and $\Sc_{_{01}}$ are  empty.
Let us  consider a numerical example with the following specifications: :
$\mu=0.8$,   $\sigma=0.3$,   $\rho=0.2$,   $\lambda=0.05$,   $\varepsilon=0.65$, and $L=0.1$. The two equations (\ref{systx13}) and (\ref{systx12})  give
$$\bar x_{_{1}} =  0.4293, \;  \bar x_{_{0-1}}=      0.9560.$$


\vspace{7mm}


\begin{small}




\end{small}

 \end{document}